\documentclass[%reprint,
twocolumn,
amsmath,
amssymb,
aps,
longbibliography,
superscriptaddress
]{revtex4-1}
\usepackage{dcolumn}% Align table columns on decimal point
\usepackage{graphicx,lineno,bm,mathtools,float}
\usepackage{epstopdf}
\begin{document}

\title{Thermodynamics of interacting single-domain superparamagnetic  nanoparticles frozen in the nodes of a regular cubic lattice}
\author{Anna Yu. Solovyova}
\author{Sergey A. Sokolsky}
\author{Ekaterina A. Elfimova}
\author{Alexey O. Ivanov}
 \email{alexey.ivanov@urfu.ru}
\affiliation{Ural Mathematical Center, Ural Federal University, \\
51 Lenin Avenue, Ekaterinburg 620000, Russia}

\date{\today}% It is always \today, today,

\begin{abstract}In this work, we study the effect of dipole-dipole interparticle interactions
on the static thermodynamic and magnetic properties of an ensemble of immobilized monodisperse
superparamagnetic  nanoparticles. We assume that magnetic nanoparticles are embedded in the nodes
of a regular cubic lattice, so that the particle translational degrees of freedom are turned off.
The relaxation of the magnetic moments of the  nanoparticles occurs by the N{\'e}el mechanism.
The easy axes are
%distributed according to the particular configuration:  these are
aligned (i) parallel or (ii) perpendicular to the direction of an external field.
These models are investigated using theory and computer simulation, taking microscopic discrete structure explicitly into account.
%The theory is based on a virial expansion of the Helmholtz free energy up to the second virial coefficient.
The analytical expressions of the Helmholtz free energy, the static magnetization, and
the initial magnetic susceptibility are derived for both configurations (i) and (ii)
as functions of the height of the magnetic crystallographic anisotropy energy barrier,
%for the internal superparamagnetic rotation of magnetic moments inside the nanoparticles
measured by parameter $\sigma$, and the intensity of the dipole-dipole interparticle
interactions measured by $\lambda_e$. A good agreement between the theory and the results of
MC simulations in the region of low and moderate values of $\lambda_e$ and $\sigma$ is obtained.
For high values of $\lambda_e$ and $\sigma$, the structuring of magnetic moments in regularly
orientated structures was found from MC simulations for configuration (i).
%A phase diagram of the orientational state of the magnetic moments on the $\lambda_e$ - $\sigma$ axes was plotted for configuration (i),  which predicts the regions of regularly orientated structuring and the absence of any structures belonging to the nanoparticle magnetic moments.
\end{abstract}

\maketitle

\section{Introduction}
Smart materials, also called responsive or  intelligent materials, are special materials that have one or more properties that can be significantly changed in a controlled fashion by external stimuli, for example a magnetic field. Such materials include magnetic composites, which are produced by embedding magnetic nanoparticles in a liquid or polymer matrix. Examples of these composites are ferrofluids, magnetic elastomers, ferrogels, magnetic emulsions, and various biocompatible magnetic fillings \cite{Kuznetsova2019,Borin2019105,Filipcsei2007,Blyakhman2019}.  Today, such materials are widely used in various medical applications because they actively respond to applied magnetic fields. For example, they are an indispensable tool in magnetic hyperthermia, in which the micromotions of magnetic nanoparticles in an alternating magnetic field leads to the heating and destruction of tumor cells \cite{Dutz2013, Ortega2013, Zubarev2018, Zubarev2019, Zhang201713929,Piehler2020,Brero20201}. The response of the magnetic composites to the applied magnetic field is determined by two main physical mechanisms of the magnetic moment's orientational relaxation in nanosized  particles. They are the Brownian rotation of particles with fixed magnetic moments and the superparamagnetic N{\'e}el rotation of magnetic moments inside the particles due to thermal fluctuations \cite{Raikher1974,Shliomis1994}. For ensembles of nanoparticles suspended in some liquid carriers, known as ferrofluids, both mechanisms take place. But when  nanoparticles are embedded in a polymer matrix  or biological tissues, the particles often lose their translational and orientational freedom. In this case, the superparamagnetic N{\'e}el relaxation becomes the major mechanism determining the magnetic properties of the ensembles of such immobilized  nanoparticles.

Today there are many established synthesis techniques that make it possible to generate a magnetic composite with various  nano- and microscopic architectures \cite{Gervald_2010, Valiev_2019,Ganesan2019315,Filipcsei2007, Weeber2018,Tanasa2019, Bastola2020377}. A different distribution of magnetic  nanoparticles inside the sample leads to a significant change in its bulk properties \cite{Zakinyan2020,Yoshida2017162,Elfimova2019}. In addition, interparticle dipole-dipole interactions have a strong effect on the macroproperties of the system, and this is manifested differently in liquid or polymer based composites. So, strong interparticle dipole-dipole interactions lead to aggregation in the ferrofluid \cite{Socoliuc2020, Elkady2015257,Pshenichnikov2012,Daffe202011222}, whereas in the system of immobilized magnetic nanoparticles, interparticle interactions can only be a cause of the structuring of the magnetic moments of the  nanoparticles, since the particles themselves remain stationary \cite{Ilg2017, Pshenichnikov2015,Solovyova2020}. The effects of dipole-dipole interactions on the bulk properties of ferrofluids are well understood theoretically \cite{Elfimova201754, Ivanov2018, Solovyova2017, Minina2018,Szalai2013}, experimentally \cite{Nagornyi2020, Lebedev2019,Linke2015,Daffe202011222}, and by computer simulation methods \cite{Minina2018, Pousaneh2020, Ivanov2018, Solovyova2017,Szalai2013,Daffe202011222}. However, devising a theory that takes into account dipole-dipole interactions in a system of immobilized superparamagnetic nanoparticles still remains a challenge.

In recent works, the magnetic response of a system of immobilized interacting single-domain  nanoparticles distributed randomly \cite{Elfimova2019} or placed at the nodes of a simple cubic lattice \cite{Solovyova2020} within an implicit solid matrix was investigated using statistical-mechanical theory and computer simulations. In the first work \cite{Elfimova2019}, superparamagnetic particles with uniaxial magnetic anisotropy were considered. The relaxation of the magnetic moments of particles occurred by the N{\'e}el mechanism. The easy axes were distributed according to the particular textures: aligned parallel or perpendicular to the external magnetic field, or randomly distributed. The initial magnetic susceptibility was found to depend on the  magnetic crystallographic anisotropy barrier (measured with respect to thermal energy by a parameter $\sigma$) in very different ways for various textures.  The initial susceptibility increased as $\sigma$ for a parallel texture, while with a perpendicular texture, the initial susceptibility decreased. With a random distribution, the initial susceptibility was independent of $\sigma$. In all cases, interactions between  nanoparticles led to an enhancement of the initial susceptibility, but the enhancement was much stronger for the parallel texture than for the perpendicular or random textures. In the second work  \cite{Solovyova2020}, it was assumed that  nanoparticles are embedded in the nodes of the simple cubic lattice (SCLF) and the  relaxation of the magnetic moments of nanoparticles occurs by the  Brownian mechanism.
The particles had no intrinsic magnetic anisotropy, but could rotate on the lattice nodes under the influence of the external magnetic field and as a result of the interparticle dipolar interactions. The magnetic properties of SCLF were compared with ones for a ferrofluid, modeled by a system of dipole hard spheres (DHS). It was found that at low intensities of the dipole-dipole interactions, the magnetization of DHS and SCLF is the same. For strong and moderate dipolar coupling regimes and with a weak magnetic field, the magnetization of the DHS system is higher than the magnetization of SCLF, while the opposite tendency is observed at stronger fields. The reasons of this behavior were discussed in the article \cite{Solovyova2020}.

These two  works \cite{Elfimova2019, Solovyova2020} demonstrated how different distributions of magnetic nanoparticles within a sample can change its macroproperties. Nevertheless, the theory is still incomplete and the topic is not fully understood. The question of how the internal magnetic anisotropy of nanoparticles affects the properties of SCLF remains unclear. The aim of this work  is to fill this gap. The properties of a monodisperse system of immobilized interacting single-domain  spherical nanoparticles, with uniaxial  magnetic  anisotropy, placed at the nodes of a simple cubic lattice will be studied in the current work using a statistical-mechanical  approach and computer simulations.

This article is arranged as follows. In Sec. \ref{sec:basis}, the theory and simulation methods are detailed, and new analytical approximations for the Helmholtz free energy are derived. The main results are presented as a comparison between theory and simulation in Sec. \ref{sec:res}. The conclusions from the work are summarized in Sec.\ref{sec:concl}.

\section{Model and methods}\label{sec:basis}
\subsection{Model}
The sample under consideration consists of $N$ immobilized superparamagnetic spherical nanoparticles distributed regularly on simple cubic lattice nodes with period~\emph{a}. All particles have the same magnetic-core diameter $x$ and magnetic moment $m = v_m M_s$, where $M_s$ is the bulk saturation magnetization,
$v_m = \pi x^3 / 6$ is the magnetic core volume. The radius vector and magnetic moment of nanoparticle $i$ are $\textbf{r}_i = r_i \hat{\textbf{r}}_i$ and $\bm{m}_i =m_i \bm{\Omega}_i$ respectively, where $\hat{\textbf{r}}_i = (\sin \theta_i \cos \phi_i, \sin \theta_i \sin \phi_i, \cos \theta_i)$ and $\bm{\Omega}_i=(\sin \omega_{i} \cos \xi_{i}, \sin \omega_{i} \sin \xi_{i}, \cos\omega_{i})$ are unit vectors.

The magnetic moment of a  nanoparticle has two degenerate ground-state directions, these being parallel and anti-parallel to the easy axis, denoted by vector $\textbf{n}_i$. The Néel energy $U_N$ as a function of the angle between $\textbf{m}_i$ and $\textbf{n}_i$ is given by
\begin{eqnarray}\label{U_N}
U_N(i) = - K v_m ( \bm{\Omega}_i \cdot \hat{\textbf{n}}_i )^2,
\end{eqnarray}
\noindent where $\hat{\textbf{n}}_i$ is a unit vector and $K$ is the magnetic crystallographic anisotropy constant (a material property).

The interaction between magnetic moment $\textbf{m}_i$ and uniform external magnetic field \textbf{H} is described by the Zeeman energy
\begin{eqnarray}\label{U_m}
U_m(i) = - \mu_0 ( \textbf{m}_i \cdot \textbf{\textrm{H}} ) = - \mu_0 m H ( \bm{\Omega}_i \cdot \hat{\textbf{\textrm{h}}} ),
\end{eqnarray}
\noindent where it is assumed that the applied magnetic field \textbf{H} has the strength $H$ and the orientation~$\hat{\textbf{\textrm{h}}}$.
\begin{figure*}[t!]
\center
\begin{minipage}[h]{65mm}
\center{\includegraphics[width=0.8\linewidth]{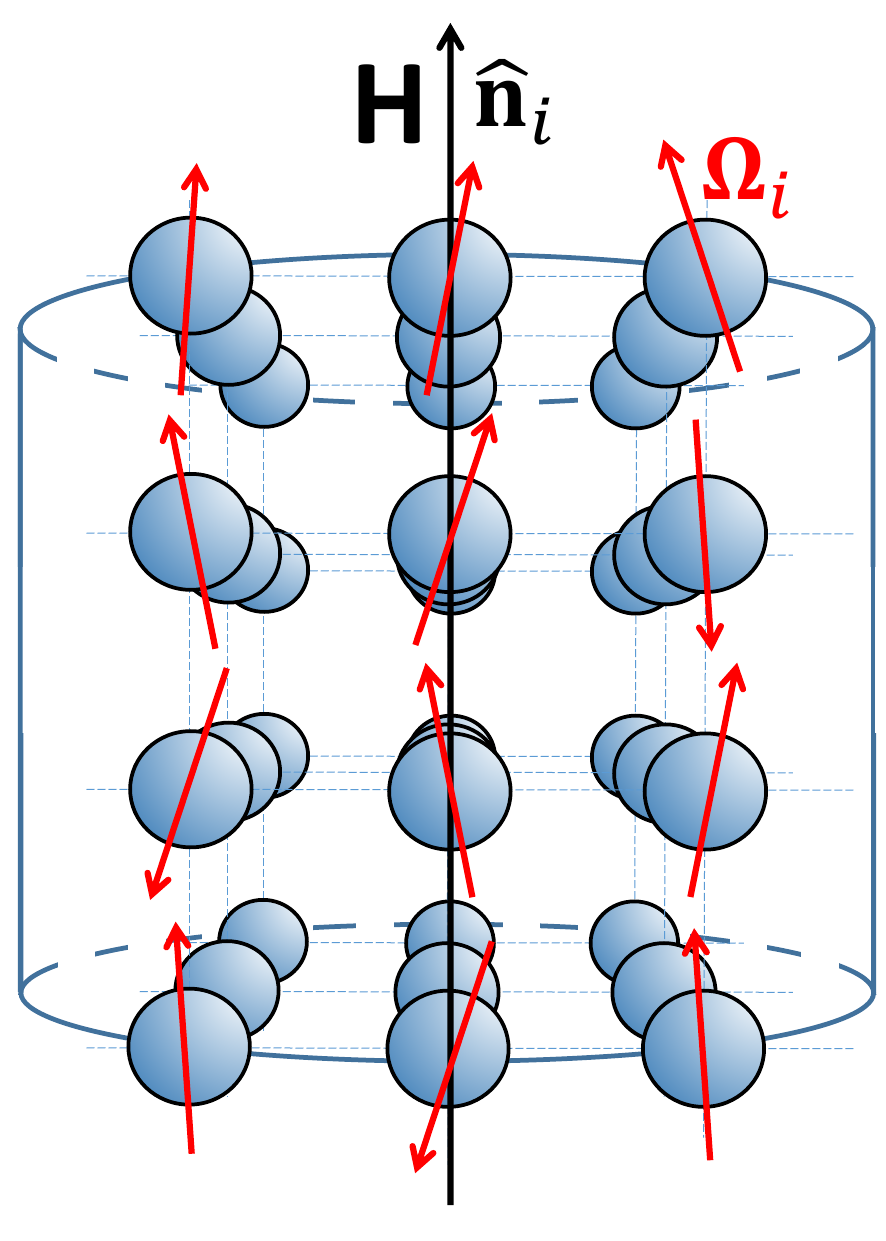}\\ (a)}
\end{minipage}
\begin{minipage}[h]{85mm}
\center{\includegraphics[width=0.8\linewidth]{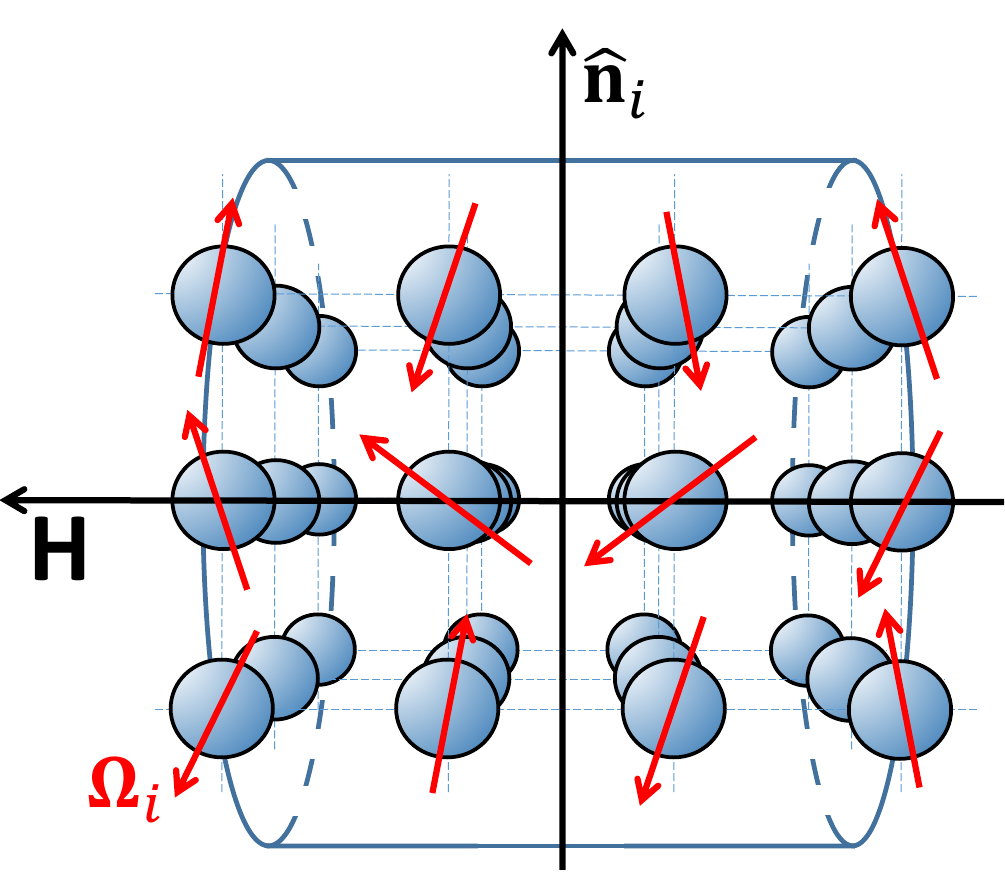}\\(b)}
\end{minipage}
\caption{Monodisperse system of immobilized single-domain superparamagnetic nanoparticles placed on the nodes of a simple cubic lattice at external magnetic field for (a) parallel and (b) perpendicular configurations.}
\label{fig:config}
\end{figure*}
The pair dipole-dipole interaction of the  two particles \emph{i} and \emph{j} obeys to the anisotropic potential $U_d$
\begin{eqnarray}\label{U_d}
U_{d}(ij) = \frac{\mu_0 m^2}{4\pi r_{ij}^3}\Big[ (\bm{\Omega}_i \cdot \bm{\Omega}_j) - 3 (\bm{\Omega}_i \cdot \hat{\textbf{r}}_{ij}) (\bm{\Omega}_j \cdot \hat{\textbf{r}}_{ij}) \Big ],
\end{eqnarray}
\noindent where $\textbf{r}_{ij} = r_{ij} \hat{\textbf{r}}_{ij}=  \textbf{r}_{j} - \textbf{r}_{i}$ has the meaning of a center-center separation vector with the length $r_{ij} = | \textbf{r}_{ij}|$.

The total potential energy normalized by the thermal energy $k_B T = \beta ^{-1}$ has the following form
\begin{eqnarray}\label{U}
\beta U &=& \beta \sum _ {j>i=1} ^ N U_{d}(ij) -\sigma \sum_{i=1}^N ( \bm{\Omega}_i \cdot \hat{\textbf{n}}_i )^2 - \alpha \sum_{i=1}^N ( \bm{\Omega}_i \cdot \hat{\textbf{h}} ) ,
\end{eqnarray}
 \noindent where the dimensionless anisotropy parameter $\sigma = \beta v_m K$ and the Langevin parameter $\alpha = \beta \mu_0 m H $ are introduced.
The relationship between the pair magnetic interaction and the thermal energy is measured by the effective dipolar coupling constant $\lambda_e$
\begin{eqnarray}\label{Lambda_e}
\lambda_e = \frac{\mu_0 m^2 \beta}{4\pi a^3},
\end{eqnarray}
\noindent the definition of which takes into account that two nanoparticles located in the nodes of the simple cubic lattice can not be closer than the lattice period $a\geq x$ .

Two types of orientational distributions of the easy axes will be considered: aligned (i) parallel and (ii) perpendicular to the direction of the external field \textbf{H}. To simplify the analytical calculations, the easy axes of both configuration are assumed to be aligned along the laboratory $Oz$ axis, namely $\hat{\textbf{n}}_i=(0,0,1)$, so the Néel energy can be represented as
\begin{eqnarray}
U_N (i) = -\sigma \cos ^2 \omega_i.
\end{eqnarray}
\noindent The direction of external magnetic field \textbf{H} is set (i) $\textbf{H} = H(0,0,1)$ in parallel configuration so that
\begin{eqnarray}
U_m^{||} (i) = -\alpha \cos \omega_i
\end{eqnarray}
\noindent and (ii) $\textbf{H} = H(1,0,0)$ in perpendicular configuration so
\begin{eqnarray}
U_m^{\perp} (i) = -\alpha \sin \omega_i \cos \xi_{i}.
\end{eqnarray}

\noindent For vanishing of demagnetization fields, we will assume that the sample container has a long cylindrical shape elongated in the direction of an external magnetic field \textbf{H}. The sample geometries studied in this work are given in Fig. \ref{fig:config}.

\subsection{The Helmholtz Free Energy of an ensemble of immobilized  nanoparticles }

\subsubsection{Ideal system}
The definition of the Helmholtz free energy $F$ contains the configurational integral $Z$ so, that
\begin{eqnarray}\label{F_def}
\beta F = - \ln \left( Z \right).
\end{eqnarray}
\noindent To extract the contribution of the dipole-dipole interactions, it is convenient to split the Helmholtz free energy $F$ into two parts
\begin{eqnarray}\label{F}
 F = F_{\textrm{\scriptsize{id}}}+ \Delta F.
\end{eqnarray}
\noindent The first term corresponds to the ideal system of superparamagnetic non-interacting nanoparticles in the applied magnetic field, the configurational integral $Z_{id}$ of which is
\begin{eqnarray}\label{dFid}
\beta  F _{\textrm{\scriptsize{id}}} &=& - \ln \left( Z_{\textrm{\scriptsize{id}}}\right),\\
Z_{\textrm{\scriptsize{id}}}&=&\prod \limits _{k=1} ^N  \int p(\textbf{r}_k) d \textbf{r}_k d\bm{\Omega}_k \nonumber \\
&\times& \exp \left( \sum \limits _{i=1}^N \big[\alpha (\bm{\Omega}_i \cdot \hat{\textbf{h}} )  + \sigma \cos ^2 \omega_i \big] \right),\label{Zid} \\
d \textbf{r}_k &=& r_k^2  \sin \theta_k d r_k d \theta_k d\phi_k ,\\
d \bm{\Omega}_k &=& \frac{1}{4 \pi}\sin \omega_k d\omega_k d\xi_k.
\end{eqnarray}
\noindent In this definition, $p(\textbf{r}_k)$
describes the probability of the location of nanoparticle \emph{k} at a given point in the volume
\begin{eqnarray}\label{PDF_solid}
p(\textbf{r}_k) = \delta (\textbf{r}_k - \textbf{r}_k^{(0)}).
\end{eqnarray}
\noindent The 'lattice position' of nanoparticle $\textbf{r}_k^{(0)}$ may obey to a regular or a random law. For both cases, the normalization rule is
\begin{eqnarray}\label{PDF_norm}
\int p(\textbf{r}_k) d \textbf{r}_k= 1.
\end{eqnarray}
\noindent Therefore in the integrand function of Eq. (\ref{Zid}), there are no dependencies on nanoparticle positions. With this integration, all nanoparticles are equivalent to each other. This allows us to rewrite $Z_{\textrm{\scriptsize{id}}}$ as follows
\begin{eqnarray}\label{Zid_}
Z_{\textrm{\scriptsize{id}}} = \left (\int d \bm{\Omega}_1 \exp \left[
\alpha (\bm{\Omega}_1 \cdot \hat{\textbf{h}} )  + \sigma \cos ^2 \omega_1
\right] \right) ^N,
\end{eqnarray}
\noindent the result of which can be calculated numerically for each set of parameters $\alpha$ and $\sigma$ and the fixed direction of vector $\hat{\textbf{n}}_1 $. It worth be stressed that the Helmholtz free energy of an ideal system

\begin{eqnarray}\label{Fid}
\frac{\beta F_{\textrm{\scriptsize{id}}}}{N} =- \ln \left (\int d \bm{\Omega}_1 \exp \left[
\alpha (\bm{\Omega}_1 \cdot \hat{\textbf{h}} )  + \sigma \cos ^2 \omega_1
\right] \right) \ \
\end{eqnarray}
does not depend on parameter $\lambda$, which characterizes the dipole-dipole interparticle interactions.

\subsubsection{A system with interparticle dipole-dipole interactions}

The second term $\Delta F$ in (\ref{F}) corresponds to the contribution of dipole-dipole correlations in the Helmholtz free energy
\begin{eqnarray}\label{dF}
\beta  \Delta F &=& - \ln \left( \frac{Z } {Z_{\textrm{\scriptsize{id}}}}\right), \\
 \frac{Z } {Z_{\textrm{\scriptsize{id}}}}&=&  \frac{1} {Z_{\textrm{\scriptsize{id}}}}\prod \limits _{k=1} ^N  \int p(\textbf{r}_k) d \textbf{r}_k d\bm{\Omega}_k \exp \Bigg( - \sum \limits _  {j>i=1} ^ N \beta U_d(ij)  \nonumber \\
&+& \sum \limits_{i=1}^N \big[\alpha (\bm{\Omega}_i \cdot \hat{\textbf{h}} )  + \sigma \cos ^2 \omega_i \big] \Bigg),\label{Z} \end{eqnarray}
\noindent  where \emph{Z} is the configurational integral of the SCLF, including dipole-dipole interaction. Introducing Boltzmann-weighted averaging over the magnetic moment orientation of nanoparticle \emph{k}
\begin{eqnarray}\label{Psi}
d \bm{\Psi}_k = \frac{ d \bm{\Omega}_k \exp \left[\alpha (\bm{\Omega}_k \cdot \hat{\textbf{h}} )  + \sigma \cos ^2 \omega_k  \right] }{\int d \bm{\Omega}_1 \exp \left[\alpha (\bm{\Omega}_1 \cdot \hat{\textbf{h}} )  + \sigma \cos ^2 \omega_1  \right]},
\end{eqnarray}
\noindent  one can rewrite definition (\ref{Z}) in a more compact form
\begin{eqnarray}\label{Znew}
\frac{Z } {Z_{\textrm{\scriptsize{id}}}}&=& \prod \limits _{k=1} ^N  \int p(\textbf{r}_k) d \textbf{r}_k d \bm{\Psi}_k \prod \limits _  {j>i=1}^N (1+f_{ij}), \\
f_{ij} &=& \exp \left(-\beta U_d(ij)\right) - 1, \label{fij}
\end{eqnarray}
\noindent where $f_{ij}$ is the Mayer function. This approach reproduces the method of \cite{Solovyova2020}, which was developed for $\sigma \equiv 0$. The main difference is the definition of the Boltzmann-weighted integration $d \bm{\Psi}_k$ (\ref{Psi}), which in our case also depends on the anisotropy parameter $\sigma \geq 0$ and coincides with the same from \cite{Solovyova2020} for $\sigma = 0$. Therefore, it is possible to apply the final results for the configurational part of the Helmholtz free energy from \cite{Solovyova2020} to our system with the new definition of operator $d \bm{\Psi}_k$:
\begin{eqnarray}\label{dF_vir}
\frac{\beta \Delta F} {N}&=& - \frac{1}{2}\sum \limits _  {j=2} ^N \left\langle f_{1j}^{(0)}\right\rangle.
\end{eqnarray}
\noindent where $f_{1j}^{(0)} = \int p(\textbf{r}_1) d \textbf{r}_1 \int p(\textbf{r}_j)d \textbf{r}_j f_{1j} $ is the Mayer function of the  nanoparticles \emph{1} and \emph{j} in their 'lattice positions'. The angle brackets in (\ref{dF_vir}) denote a Boltzmann-weighted integration (\ref{Psi}) over the orientation of both  nanoparticles \emph{1} and \emph{j}
\begin{eqnarray}
\left\langle f_{1j}^{(0)} \right\rangle = \int f_{1j}^{(0)} d \bm{\Psi}_1 d \bm{\Psi}_j  .
\end{eqnarray}
\noindent It should be emphasized that approximation (\ref{dF_vir}) is obtained due to the logarithm expansion of (\ref{dF}) up to the linear term, and it is limited by the second virial coefficient level, which means the consideration of the interparticle interactions in nanoparticle pairs only.

This definition of $\Delta F$ is valid for the regular types of  nanoparticle distribution in the system volume, such as SCLF or other lattices. For the random distribution of magnetic nanoparticles in the sample volume, it is necessary to average (\ref{dF_vir}) over all possible random configurations. This means that in the limit, every particle $j$ in sum of (\ref{dF_vir}) can occupy any position in the sample volume except for that of particle~1:
\begin{eqnarray}
\frac{\beta \Delta F ^{\textrm{random}}} {N} &=& - \frac{1}{2}\sum \limits _  {j=2} ^N  \int \frac{dr_{1j}}{V} \left\langle f_{1j}\right\rangle \\
&=& - \frac{N}{2V}  \int dr_{12}\left\langle f_{12}\right\rangle =  \rho B_2, \nonumber
\end{eqnarray}
\noindent where $\rho=N/V$ is nanoparticle density, $B_2$ corresponds the classical definition of the second virial coefficient in the theory of liquids \cite{Balescu1975}.

Next we will consider in detail the behavior of SCLF.
The Mayer function (\ref{fij}) is expanded  in a series up to the third order in terms of the dipolar energy $U_d$
\begin{eqnarray}
f_{1j}^{(0)} = -\beta U_d (1j) + \frac{\left[ -\beta U_d(1j) \right]^2}{2!}  + \frac{\left[ -\beta U_d (1j) \right]^3}{3!} . \ \ \ \
\end{eqnarray}
\noindent Dipole-dipole potential $U_d$ includes the dependencies over both the translational and orientational degrees of freedom of nanoparticles 1 and \emph{j}. Applying the Boltzmann-weighted integration over the magnetic moment orientations $\textbf{m}_{1}$ and $\textbf{m}_{j}$, it is possible to represent the value of $\beta \Delta F$ in the common form
\begin{eqnarray}\label{b123}
\frac{\beta \Delta F} {N} &=& - \frac{1}{2} \left(b_1 \lambda_e + b_2 \lambda_e^2 + b_3 \lambda_e^3 \right),\\
b_1 &=&\sum \limits _  {j=2} ^N \bigg\langle \frac{-\beta U_d (1j)}{\lambda_e}\bigg\rangle, \\
b_2 &=& \sum \limits _  {j=2} ^N \bigg\langle \frac{1}{2!} \left(\frac{ -\beta U_d(1j)}{\lambda_e} \right)^2\bigg\rangle, \\
b_3 &=&  \sum \limits _  {j=2} ^N \bigg\langle \frac{1}{3!} \left(\frac{ -\beta U_d(1j)}{\lambda_e} \right)^3\bigg\rangle.
\end{eqnarray}
\noindent The coefficients $b_1$, $b_2$ and $b_3$ are different for the parallel and perpendicular configurations and will be discussed separately.

\subsection{The parallel configuration of SCLF with superparamagnetic nanoparticles}
\begin{figure}[b!]
\center
\includegraphics[width=0.95\linewidth]{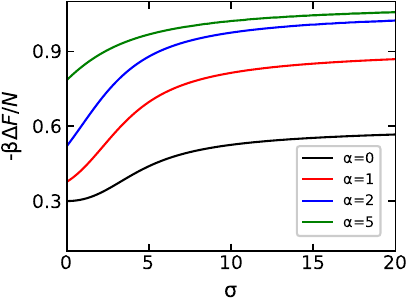}
\caption{The contribution of dipole-dipole interactions $\Delta F$ for the parallel configuration as a function of the anisotropy parameter $\sigma$ for the system with $\lambda_e = 0.5$ and different values of $\alpha$ = 0, 1, 2 , and 5. The value of $\alpha$ increases from the bottom to the top.  }
\label{fig:Fs}
\end{figure}

The parallel configuration corresponds to Fig.~\ref{fig:config}~(a), which means $( \bm{\Omega}_i \cdot \hat{\textbf{h}} ) = ( \bm{\Omega}_i \cdot \hat{\textbf{n}}_i ) = \cos \omega _i$. In this case, the ideal part of the Helmholtz free energy (\ref{Fid}) is equal to
\begin{eqnarray}\label{Fid_para}
\frac{\beta F_{\textrm{\scriptsize{id}}}}{N} =- \ln \left[ Q_0(\alpha,\sigma) \right],
\end{eqnarray}
\noindent where
\begin{eqnarray}
Q_0(\alpha,\sigma) &=& \frac{1}{2} \int \limits_{-1}^1 \exp \left(\alpha t  + \sigma t ^2  \right) d t,\\
Q_0(\alpha, 0) &=& \frac{\sinh{\alpha}}{\alpha}.
\end{eqnarray}
The Boltzmann-weighted integration over the orientation of the nanoparticle magnetic moments for the parallel configuration is
\begin{eqnarray}
d \bm{\Psi}_k = \frac{ d \bm{\Omega}_k \exp \left(\alpha \cos \omega_k  + \sigma \cos ^2 \omega_k  \right) }{Q_0(\alpha,\sigma)}.
\end{eqnarray}
\noindent Details of the averaging of coefficients $b_1$, $b_2$, $b_3$ over magnetic moment orientations and particle positions are provided in  Appendix A. The final analytical expression for $\Delta F$ expansion can be presented as
\begin{eqnarray}\label{dF_vir_par}
\frac{\beta \Delta F} {N} &=& -2.0944 Q_1^2(\alpha,\sigma) \lambda_e \nonumber \\
&-& \left[1.9390 Q_2^2(\alpha,\sigma)+1.4003\right] \lambda_e^2 \nonumber \\
&-& \big[0.1611 Q_3^2 (\alpha,\sigma) + 1.5595 Q_1(\alpha,\sigma) Q_3(\alpha,\sigma) \nonumber \\
&-& 0.7285 Q_1^2(\alpha,\sigma) \big] \lambda_e^3 ,
 \end{eqnarray}
\noindent where additional functions $Q_k(\alpha,\sigma)$ are defined in Appendix A.
It is well known that the virial expansion for a system with dipole-dipole interactions is an alternating series \cite{Henderson2011,Joslin19811507}, and therefore it is very sensitive to truncation of the series. Hence, it could be more efficient to transform the virial expansion of the Helmholtz free energy (\ref{dF_vir_par}) back into the logarithmic form (\ref{dF}):
\begin{eqnarray}\label{dF_log_par}
\frac{\beta \Delta F } {N}&=& - \ln \Big\{ 1+ 2.0944 Q_1^2(\alpha,\sigma) \lambda_e  \nonumber \\
&+& \left[1.9390 Q_2^2(\alpha,\sigma)+1.4003\right] \lambda_e^2 \nonumber \\
&+& \big[0.1611 Q_3^2 (\alpha,\sigma) + 1.5595 Q_1(\alpha,\sigma) Q_3(\alpha,\sigma)  \nonumber \\
&-& 0.7285 Q_1^2(\alpha,\sigma) \big] \lambda_e^3 \Big\} ,
 \end{eqnarray}
\noindent so that the terms from the right-hand side of Eq. (\ref{dF_vir_par}) are the first terms of the Maclaurin expansion of the logarithm in (\ref{dF_log_par}). The advantage of the logarithmic form is that the logarithm of a polynomial is less sensitive to polynomial truncation at the low order in $\lambda_e$-expansion. First suggested in \cite{Elfimova2012}, this method allows for the expansion of the theory applicability over both the volume nanoparticle concentration and the dipolar coupling constant for the prediction of the properties of the dipolar hard sphere fluid \cite{Elfimova2013,Solovyova2020,Solovyova2020_poly}. The typical behavior of the dipole-dipole contribution in the Helmholtz free energy $\Delta F$ as a function of the anisotropy parameter $\sigma$ is shown in Fig. \ref{fig:Fs} with $\alpha = 0$, 1, 2, and 5, and a rather high value of intensity of dipole-dipole interactions $\lambda_e = 0.5$. For $0\leq \sigma \leq  5$, one can note that the contribution $\Delta F$ increases more rapidly than for $\sigma >5$.
\begin{figure}[b!]
\center
\includegraphics[width=0.95\linewidth]{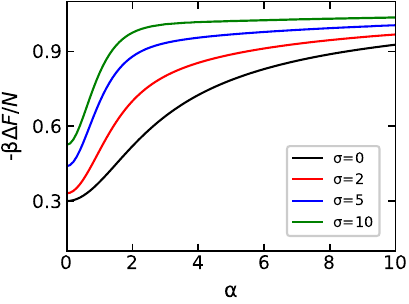}
\caption{The contribution of dipole-dipole interactions $\Delta F$ for the parallel configuration as a function of the Langevin parameter $\alpha$ for the system with $\lambda_e = 0.5$ and different values of $\sigma$ = 0, 2, 5 and 10. The value of $\sigma$ increases from the bottom to the top.}
\label{fig:Fa}
\end{figure}
In the parallel configuration, anisotropy axes are an addition stimulus to the alignment of magnetic moments along the external magnetic field, which leads an increase in the Helmholtz free energy as the anisotropy parameter $\sigma$. The dependence on $\Delta F$ from the intensity of the external magnetic field is shown in Fig. \ref{fig:Fa}, with $\sigma = 0$, 2, 5, and 10, and $\lambda_e = 0.5$. The contribution $\Delta F$ is more sensitive to changes in the  anisotropy parameter in the intermediate magnetic fields $1\leq \alpha\leq 5$, whereas at $\alpha \simeq 10$ the difference between values of $\Delta F$ as $\sigma$ increases is not sensitive.

\subsection{The perpendicular configuration of SCLF with superparamagnetic nanoparticles}

The perpendicular configuration is illustrated in Fig. \ref{fig:config} (b), which corresponds to $( \bm{\Omega}_i \cdot \hat{\textbf{h}} ) = \sin \omega_{i} \cos \xi_{i}$. In this case, the ideal part of the Helmholtz free energy (\ref{Fid}) is equal to
\begin{eqnarray}\label{Fid_perp}
\frac{\beta F_{\textrm{\scriptsize{id}}}}{N} =- \ln \left [R_0(\alpha,\sigma) \right],
\end{eqnarray}
\noindent where
\begin{eqnarray}
R_0(\alpha,\sigma) &=& \int \limits_{0}^1 \exp \left(\sigma t ^2  \right) I_0 (\alpha \sqrt{1-t^2})d t, \\
R_0(\alpha,0) &=& \frac{\sinh{\alpha}}{\alpha}.
\end{eqnarray}
\noindent  Here $I_0(x)$ is the modified Bessel function of zero order.

The Boltzmann-weighted integration over the orientation of the  nanoparticle magnetic moment for the perpendicular configuration is
\begin{eqnarray}
d \bm{\Psi}_k = \frac{ d \bm{\Omega}_k \exp \left(\alpha \sin \omega_{k} \cos \xi_{k}  + \sigma \cos ^2 \omega_k  \right) }{R_0(\alpha,\sigma)}.
\end{eqnarray}
\noindent Details of the averaging of the coefficients $b_1$, $b_2$, $b_3$ over magnetic moment orientations and particle positions are provided in  Appendix B.  The final analytical expression for $\Delta F$ in the logarithmic form is
\begin{eqnarray}\label{dF_log_perp}
\frac{\beta \Delta F } {N}&=& - \ln \Big\{ 1+ 2.0944 R_1^2(\alpha,\sigma) \lambda_e \nonumber \\
&+& \big[ 1.4542 R_2^2 (\alpha,\sigma) + 4.3627 R_3^2 (\alpha,\sigma) \nonumber \\
&-& 2.9085 R_3 (\alpha,\sigma) + 1.8851 \big] \lambda_e^2 \nonumber \\
&+& \big[ 2.8736 R_4 (\alpha,\sigma) R_5 (\alpha,\sigma)- 0.1429 R_4 (\alpha,\sigma) R_6 (\alpha,\sigma)  \nonumber \\
&+& 0.2454 R_5 (\alpha,\sigma) R_6 (\alpha,\sigma) - 1.3856 R_4^2 (\alpha,\sigma) \nonumber \\
&-&0.4960 R_5^2 (\alpha,\sigma)   - 1.3856 R_6^2 (\alpha,\sigma) \big] \lambda_e^3 \Big\} .
 \end{eqnarray}
\noindent Functions $R_k(\alpha,\sigma)$ are defined in Appendix B.

\begin{figure}[h!]
\center
\includegraphics[width=\linewidth]{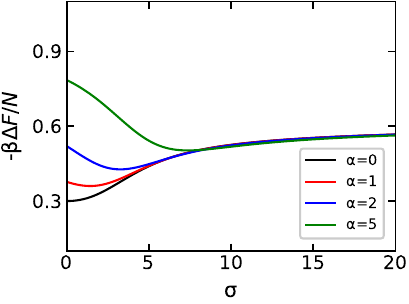}
\caption{The contribution of dipole-dipole interactions $\Delta F$ for the perpendicular configuration as a function of the anisotropy parameter $\sigma$ for the system with $\lambda_e = 0.5$ and different values of $\alpha$ = 0, 1, 2 and 5.}
\label{fig:Fs_perp}
\end{figure}

\begin{figure}[h!]
\center
\includegraphics[width=\linewidth]{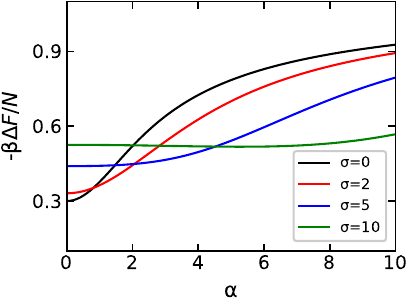}
\caption{The contribution of dipole-dipole interactions $\Delta F$ for the perpendicular configuration as a function of the Langevin parameter $\alpha$ for the system with $\lambda_e = 0.5$ and different values of $\sigma$ = 0, 2, 5 and 10.}
\label{fig:Fa_perp}
\end{figure}

The dependence of the dipole-dipole contribution in the Helmholtz free energy $\Delta F$ as a function of the anisotropy parameter $\sigma$ is shown in Fig. \ref{fig:Fs_perp} with $\alpha = 0$, 1, 2, and 5, and $\lambda_e = 0.5$. It should be noted that these dependencies are not monotonic except for $\alpha = 0$ (black curve). In the low values of the anisotropy parameter, the increase of magnetic field intensity leads to an increase in the dipole-dipole contribution to the Helmholtz free energy, but than all curves coincide with each other and show a constant value for $\sigma \geq  8$. This means that for the perpendicular configuration, the dipolar part of the Helmholtz free energy $\Delta F$ for the model system with $\sigma \gg 0$ depends mainly on the intensity of dipolar interaction $\lambda$ and is almost independent of $\alpha$ and $\sigma$. This fact can also be seen in Fig. \ref{fig:Fa_perp}, where $\Delta F$ is shown as a function of $\alpha$. At $\sigma = 10$ (green curve), the behavior of $\Delta F$ is not very sensitive to increases in the Langevin parameter.

\subsection{Simulation}
In order to check the accuracy of the new theory and discover its applicability, its predictions were thoroughly tested  against computer simulation data. Monte Carlo (MC) simulations were carried out in a canonical (\emph{NVT}) ensemble for $N=512$ dipolar hard spheres embedded  in the lattice nodes. The model configuration was generated inside a cubic box, to which 3D periodic boundary conditions were applied. To vanish all the demagnetization effects, the Ewald summation with conducting boundary conditions was used for computing the long-range dipole-dipole interactions between magnetic nanoparticles. It was assumed that the external magnetic field was directed along the $Oz$ axis. The easy axes were the unit vectors parallel to (i) the laboratory \emph{Oz} axis in the parallel configuration and (ii) the laboratory \emph{Ox} axis in the perpendicular configuration. To overcome the anisotropy barrier for high values of $\sigma$, there were two equiprobable types of rotational move: ordinary random displacement and the flip move $\textbf{m} \rightarrow - \textbf{m}$ \cite{Elfimova2019}. Typical run lengths consisted of $10^6$ attempted rotations per  nanoparticle after equilibration. Estimates of statistical errors were calculated using the blocking procedure described in \cite{Allen1987}. In all cases, the obtained values of statistical errors do not exceed the size of symbols used for the simulation data.

\begin{figure}[h!]
\center
\includegraphics[width=\linewidth]{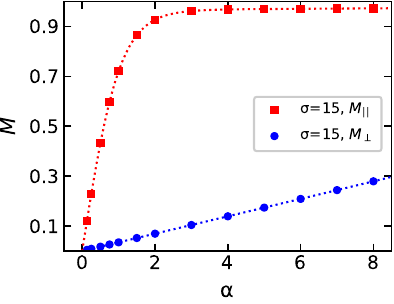}
\caption{The static magnetization as a function of the Langevin parameter $\alpha$ for a system with $\sigma=15$ in the absence of interparticle interactions. Dashed lines correspond to the ideal approximation (\ref{mag_id}). The symbols are from Monte-Carlo simulations. The results are shown for the parallel (red squares and line) and perpendicular (blue circles and line) configurations.}
\label{fig:mag_mc_id}
\end{figure}

\begin{figure}[h!]
\center
\includegraphics[width=\linewidth]{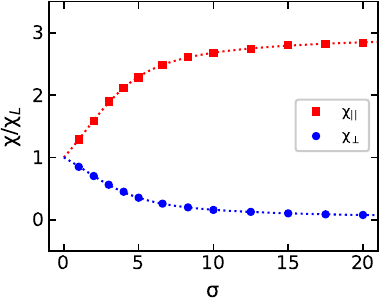}
\caption{The initial magnetic susceptibility $\chi$ divided by the Langevin susceptibility $\chi_L$ as a function of the anisotropy parameter $\sigma$ for a system without interparticle interactions. Dashed lines correspond to the ideal approximation (\ref{chi_id}). The symbols are from Monte-Carlo simulations. The results are shown for the parallel (red squares and lines) and perpendicular (blue circles and lines) configurations.}
\label{fig:chi_mc_id}
\end{figure}

The Helmholtz free energy can not be measured by the numerical method directly, so its derivatives (the scalar magnetization and the initial magnetic susceptibility) were used to investigate the validity of the new theory. For both the parallel and perpendicular configurations, fractional magnetization was computed in the simulation as
\begin{eqnarray}\label{sim_magn}
M = \frac{1}{N}\left\langle \sum_{i = 1}^{N} \cos \omega_i \right\rangle_{t},
\end{eqnarray}
\noindent where $\langle \ldots \rangle_t$ means the average over simulation time. The initial magnetic susceptibility was calculated at $\alpha = 0$ in the $z$ direction only:
\begin{eqnarray}\label{sim_susc}
\chi = \chi_L \Bigg\langle \left(\sum_{i = 1}^{N} \cos \omega_i \right)^2 \Bigg\rangle_{t} \frac{3}{N},
\end{eqnarray}
\noindent where the Langevin susceptibility $\chi_L$ can be expressed via parameter $\lambda_e$ using the relation $\rho = 1/a^3$ for the simple cubic lattice
\begin{eqnarray}\label{chi_L}
\chi_L = \frac{4 \pi \mu_0 \rho m^2}{3 k_B T} = \frac{4\pi}{3} \lambda_e.
\end{eqnarray}
To check the simulation algorithm with high values of the anisotropy parameter, an ideal system of superparamagnetic nanoparticles was modeled where interparticle interaction was turned off. In this case, the exact theoretical results are known:
\begin{eqnarray}\label{mag_id}
M_{\textrm{\scriptsize{id}}} &=& - \frac{\partial}{\partial \alpha} \left( \frac{\beta F_{\textrm{\scriptsize{id}}}}{N} \right), \\
\chi_{\textrm{\scriptsize{id}}} &=&- \frac{1}{V} \frac{\partial^2  F_{\textrm{\scriptsize{id}}}}{\partial H^2}\Bigg | _{H=0},\label{chi_id}
\end{eqnarray}
\noindent where $F_{\textrm{\scriptsize{id}}}$ is defined by (\ref{Fid_para}) for the parallel configuration and (\ref{Fid_perp}) for the perpendicular configuration. In Figs. \ref{fig:mag_mc_id} and~\ref{fig:chi_mc_id}, one can find the excellent agreement between obtained simulation data and ideal approximations (\ref{mag_id}) and (\ref{chi_id}) for both magnetization and the initial magnetic susceptibility.


\begin{figure*}[t!]
\center
\begin{minipage}[h]{110mm}
\center{\includegraphics[width=\linewidth]{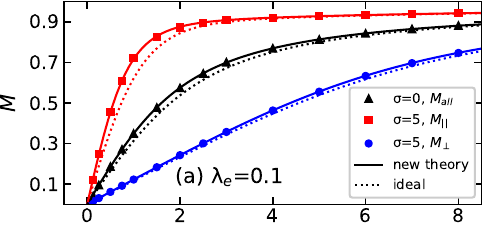}}
\end{minipage}
\begin{minipage}[h]{110mm}
\center{\includegraphics[width=\linewidth]{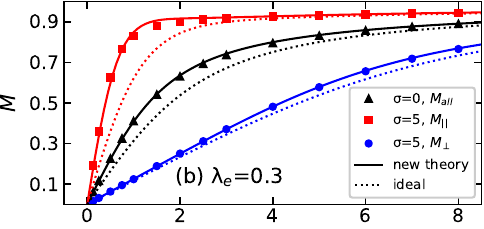}}
\end{minipage}
\begin{minipage}[h]{110mm}
\center{\includegraphics[width=\linewidth]{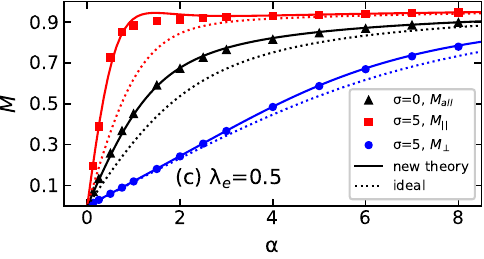}}
\end{minipage}
\caption{The static magnetization as a function of the Langevin parameter $\alpha$ for systems with (a) $\lambda_e = 0.1$, (b) 0.3 and (c) 0.5. The solid lines correspond to the new theories for $M_{||}$ and $M_{\perp}$, the dashed lines are the ideal approximation (\ref{mag_id}). The symbols are from Monte-Carlo simulations. The results are shown for the parallel (red squares and lines) and the perpendicular (blue circles and lines) configurations with $\sigma=5$ and for the system with $\sigma = 0$ (black triangles and lines).}
\label{fig:mag_mc}
\end{figure*}
\section{Results}\label{sec:res}

The analytical expression of the Helmholtz free energy allows us to obtain predictions for the various magnetic and thermodynamic properties of the system.
So, the scalar magnetization  is defined by
\begin{eqnarray}\label{mag}
 M &=&  M_{\textrm{\scriptsize{id}}} - \frac{\partial}{\partial \alpha} \left( \frac{\beta \Delta F}{N} \right),
\end{eqnarray}
\begin{figure*}[t!]
\center
\begin{minipage}[h]{65mm}
\center{\includegraphics[width=\linewidth]{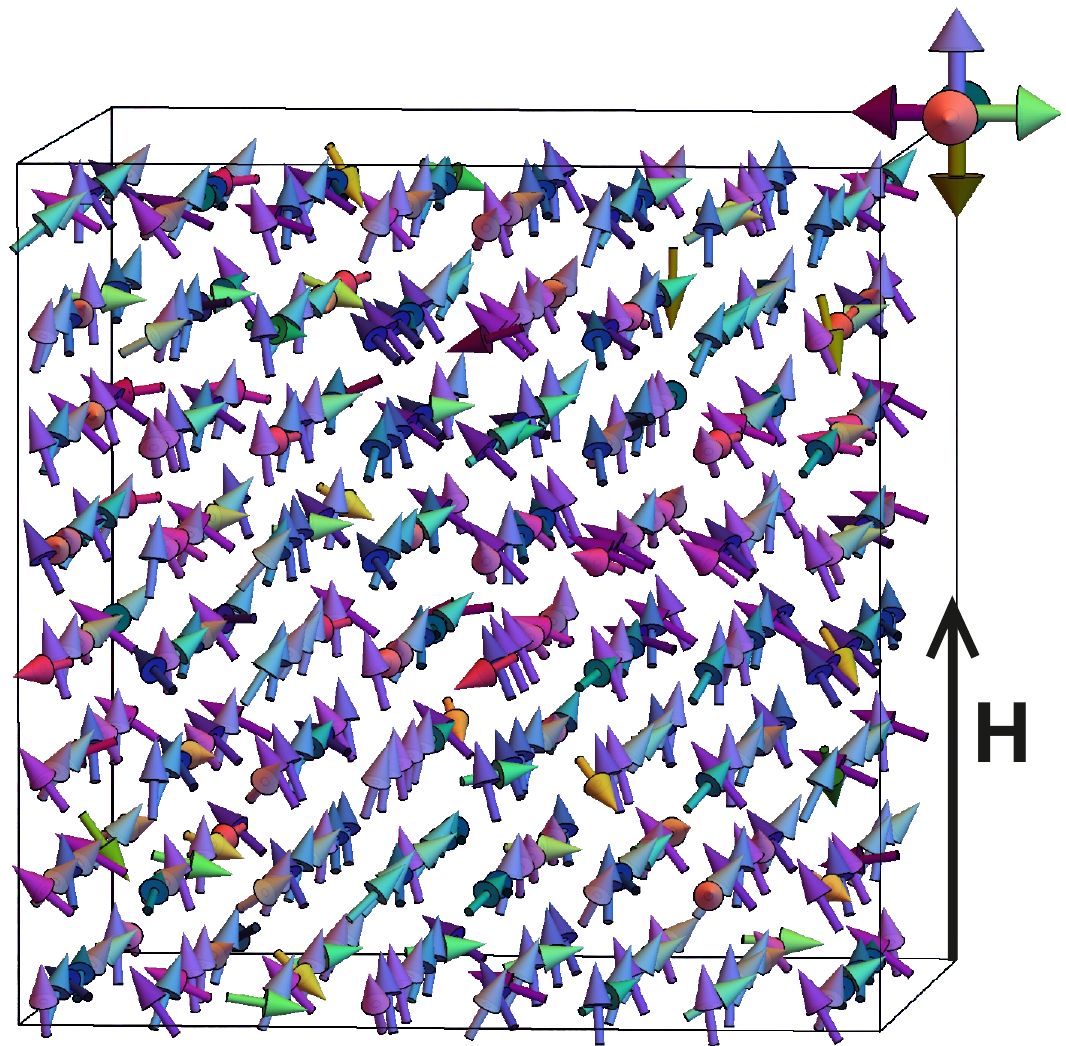} \\ (a) $\alpha=1$, $\sigma = 0$, $M_{all} = 0.45$}
\end{minipage}
\begin{minipage}[h]{65mm}
\center{\includegraphics[width=\linewidth]{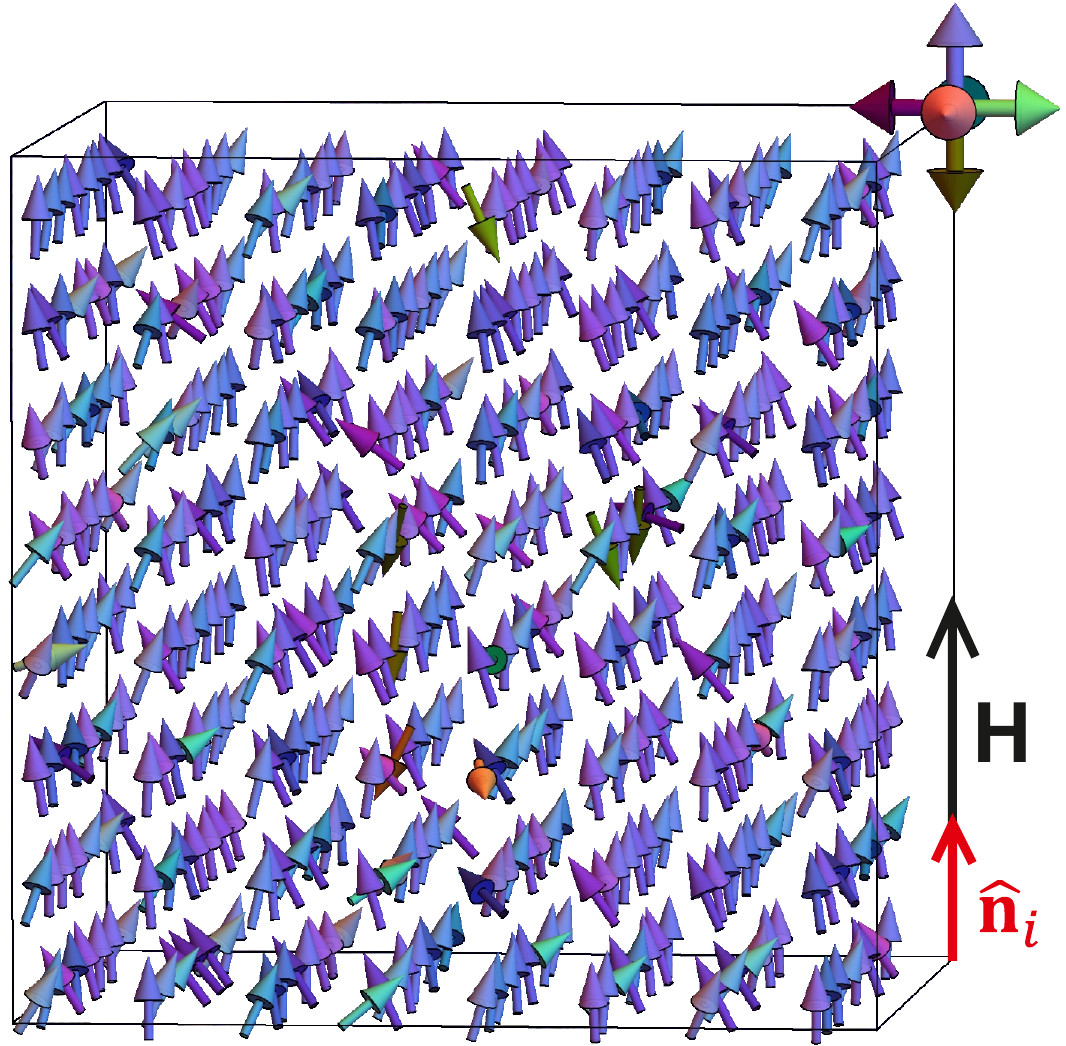}\\(b) $\alpha=1$, $\sigma = 5$, $M_{\|} = 0.88$}
\end{minipage}
\begin{minipage}[h]{65mm}
\center{\includegraphics[width=\linewidth]{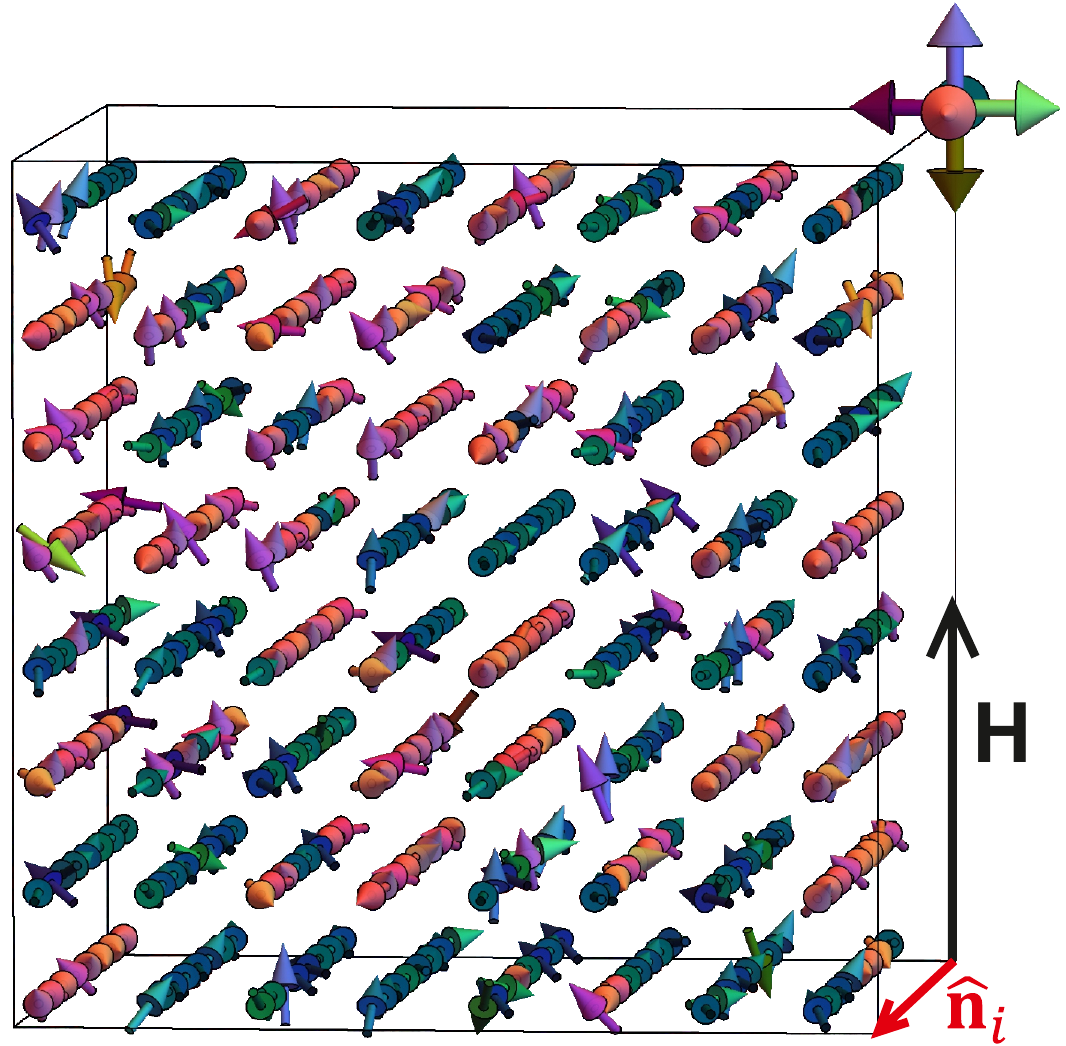}\\(c) $\alpha=1$, $\sigma = 5$, $M_{\perp} = 0.12$}
\end{minipage}
\caption{Simulation snapshots of a model system with $\lambda_e=0.5$ and
$\alpha = 1$. The results are shown for (a) $\sigma = 0$, (b) parallel configurations with $\sigma=5$ and (c) perpendicular configurations with $\sigma=~5$. Different arrow colors correspond to different orientations of the magnetic moments.}
\label{fig:mag_snapshots}
\end{figure*}
\noindent where $\Delta F$ can be found in (\ref{dF_log_par}) for the parallel configuration and (\ref{dF_log_perp}) for the perpendicular configuration. The second term takes into account the interparticle dipole-dipole interactions in the model system. Three systems were considered: $\lambda_e =0.1$, 0.3 and 0.5. The Langevin susceptibilities for these systems are equal to $\chi_L=0.42$, 1.26 and 2.1, respectively. In Fig. \ref{fig:mag_mc}, the theoretical static magnetization curves are compared with the MC simulation data. For $\sigma = 0$ (black curves), the theoretical predictions for both the parallel and perpendicular configurations coincide with each other, as was expected. In this case, the excellent agreement of the new theory (solid lines) with the simulation data for all considered systems should be noted, whereas the ideal approximation (dashed lines) works well only for a system with a weak interactions ($\lambda_e = 0.1$). The same can be said about the perpendicular configuration with $\sigma = 5$ (blue curves). The magnetization of the parallel configuration with $\sigma=5$ increases rapidly as the magnetic field intensity increases, and the dipole-dipole interaction effect is more pronounced in this case. However, for $\sigma = 5$ and $\lambda_e=0.5$, a small deviation between the new theoretical formula (\ref{mag}) and the simulation data is observed in the range of magnetic field $1 \leq  \alpha \leq 2$.

The typical behavior of the magnetization can be described as follows:  an increase in $\sigma$ leads to an increase in the magnetization in the parallel case and a decrease in the perpendicular configuration. This fact is clearly illustrated in Fig. \ref{fig:mag_snapshots}, where simulation snapshots are given for the model system  with $\lambda_e=0.5$ and $\alpha = 1$. Fig. \ref{fig:mag_snapshots} (a) corresponds to when $\sigma = 0$ and scalar magnetization is equal to $M = 0.45$. The presence of an easy axis with $\sigma = 5$ along the external magnetic field almost doubles the scalar magnetization, as shown in Fig. \ref{fig:mag_snapshots} (b). The perpendicular configuration with $\sigma = 5$ is given in Fig. \ref{fig:mag_snapshots} (c), where the system's magnetic response to the external magnetic field is very weak. In this case, the magnetic moments formed chains in the direction of the easy axis so that scalar magnetization in the direction of $\hat{\textbf{n}}_i$ is equal to zero, although a pronounced regular structure in the magnetic moments of all the system's nanoparticles is not yet observed.
\begin{figure*}[t!]
\center
\begin{minipage}[h]{110mm}
\center{\includegraphics[width=\linewidth]{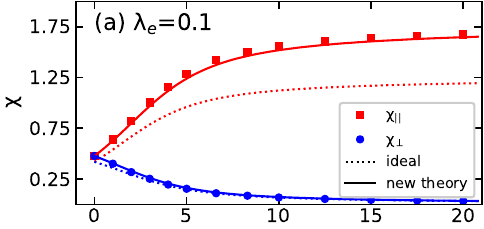}}
\end{minipage}
\begin{minipage}[h]{110mm}
\center{\includegraphics[width=\linewidth]{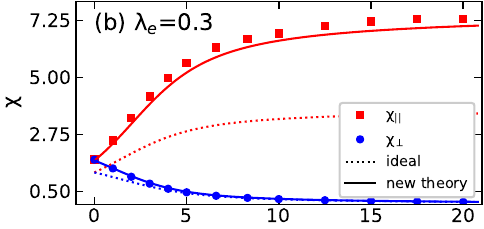}}
\end{minipage}
\begin{minipage}[h]{110mm}
\center{\includegraphics[width=\linewidth]{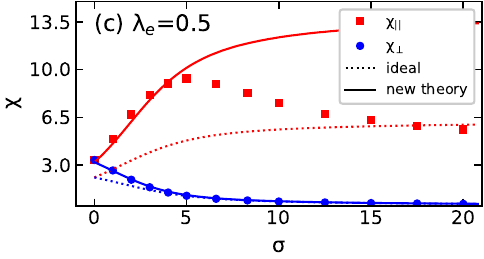}}
\end{minipage}
\caption{The initial magnetic susceptibility $\chi$ as a function of anisotropy parameter $\sigma$ for systems with (a) $\lambda_e = 0.1$, (b) 0.3 and (c) 0.5. The solid lines correspond to new theories for $\chi_{||}$ and $\chi_{\perp}$, the dashed lines are the ideal approximation (\ref{chi_id}). The symbols are from Monte-Carlo simulations. The results are shown for the parallel (red squares and lines) and perpendicular (blue circles and lines) configurations.}
\label{fig:chi_mc}
\end{figure*}

The initial slope of the magnetization curve is characterized by the initial magnetic susceptibility, which can be determined via the Helmholtz free energy as
\begin{eqnarray}\label{chi}
 \chi = \chi_{\textrm{\scriptsize{id}}} - \frac{1}{V} \frac{\partial^2 \Delta F}{\partial H^2}\Bigg | _{H=0}.
\end{eqnarray}
\noindent A comparison between the theoretical predictions of $\chi$ and the MC results is given in Fig. \ref{fig:chi_mc}. The results are shown for both the parallel and perpendicular configurations. As for ideal approximations (\ref{chi_id}), there is a significant discrepancy between the dashed curves and the simulation data in all the considered systems, but especially for the parallel configuration. Even for a weakly interacting system with $\lambda_e=0.1$, see Fig. \ref{fig:chi_mc} (a), the difference between the susceptibilities of the non-interacting (red dashed line) and interacting  nanoparticles (red solid line) for the parallel configuration is surprisingly large. Interactions lead a rapid growth in susceptibility, and the new theory (\ref{chi}) allows for an accurate description of this behavior in $\chi_{\|}$. One can note a strong agreement between the new theory (solid lines) and the MC data (symbols) for both the parallel and perpendicular configurations at $\lambda_e=0.1$. As $\lambda_e$ increases up to 0.3 in Fig. \ref{fig:chi_mc} (b), a small deviation between the new theory (\ref{chi}) and the simulation data appears in the parallel configuration (red color), while for the perpendicular configuration (blue color), the agreement of theoretical and numerical results remains strong. The MC results for the parallel configuration at $\lambda_e=0.5$ (Fig. \ref{fig:chi_mc} (c)) demonstrate an unexpected  effect: the non-monotonic behavior of susceptibility as $\sigma$ increases. Note that the maximum achievable susceptibility value in this case is $\chi_{\|}\simeq 9$ at $\sigma \simeq 5$.

\begin{figure}[t!]
\center
\begin{minipage}[h]{67mm}
\center{\includegraphics[width=\linewidth]{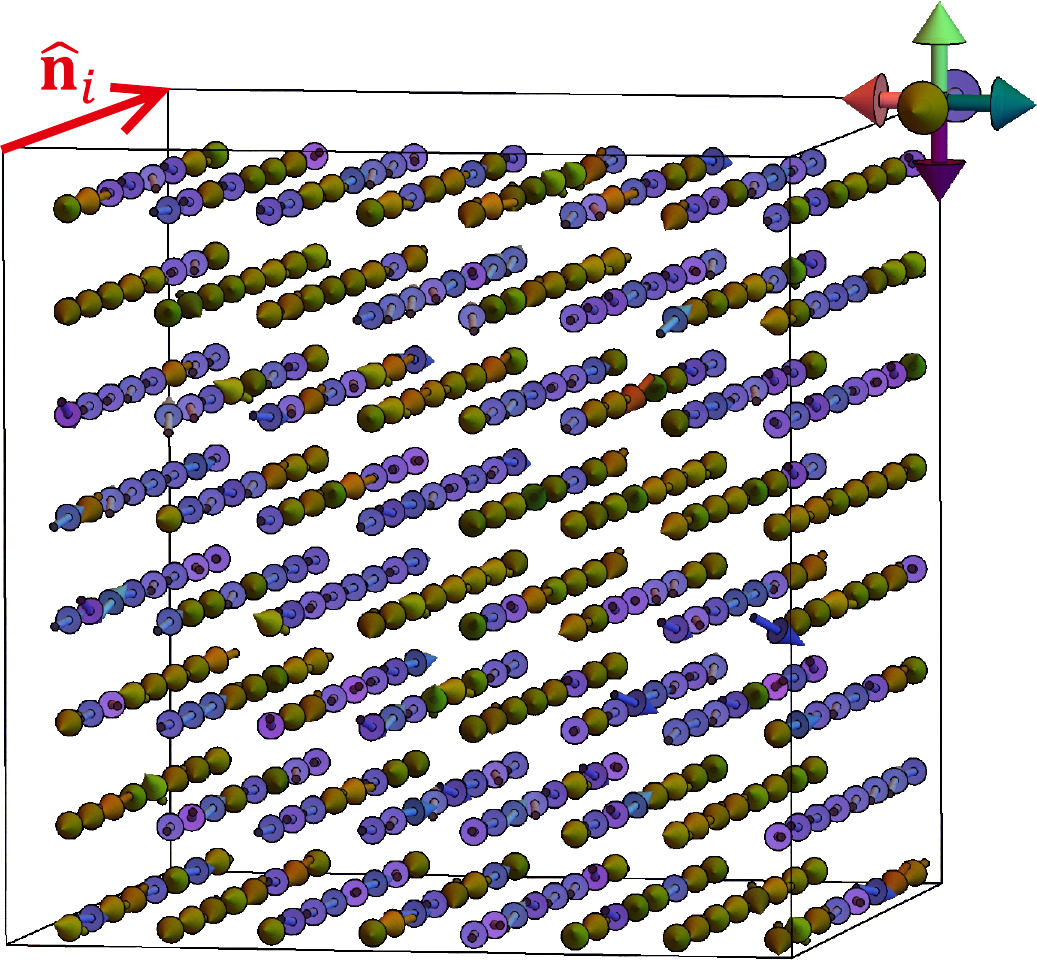} \\ (a) $\lambda_e = 0.3$, $\chi_{\|} = 7.32$}
\end{minipage}
\begin{minipage}[h]{67mm}
\center{\includegraphics[width=\linewidth]{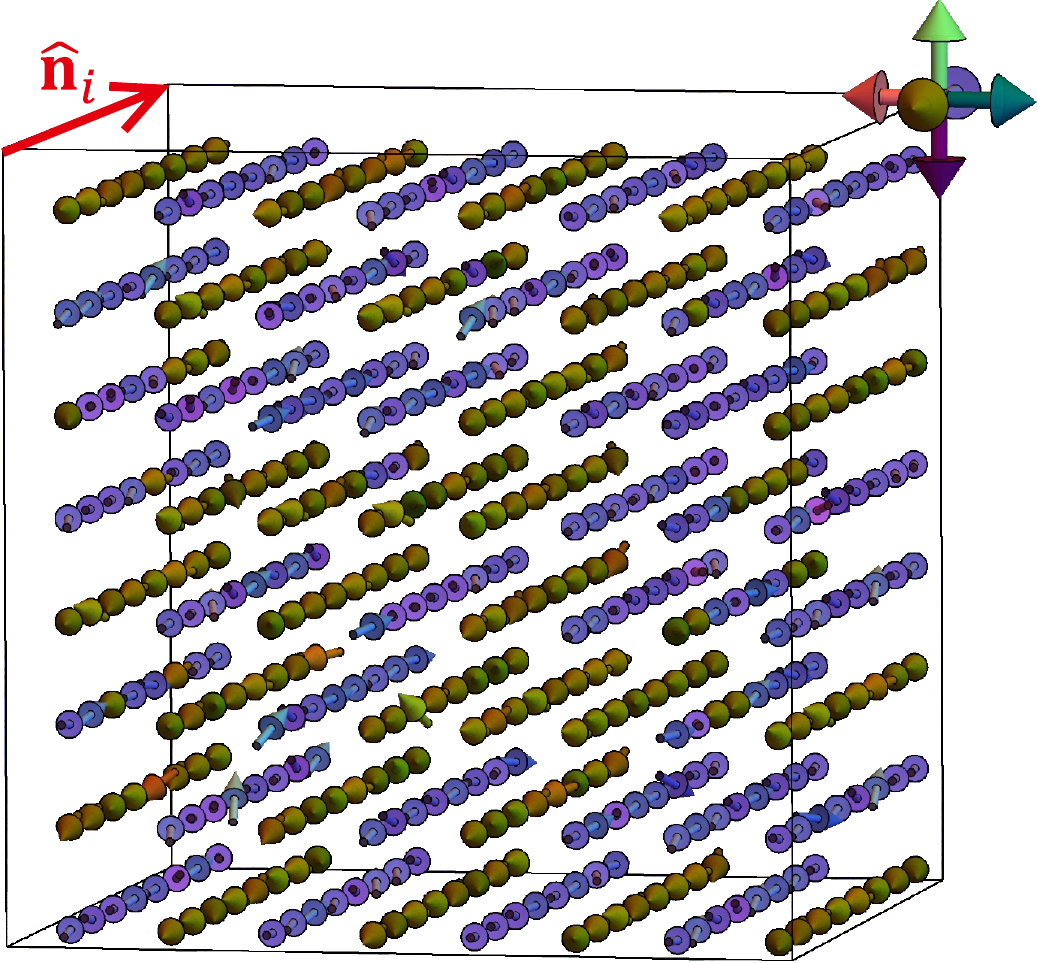}\\(b) $\lambda_e = 0.5$,  $\chi_{\|} = 5.62$}
\end{minipage}
\begin{minipage}[h]{67mm}
\center{\includegraphics[width=\linewidth]{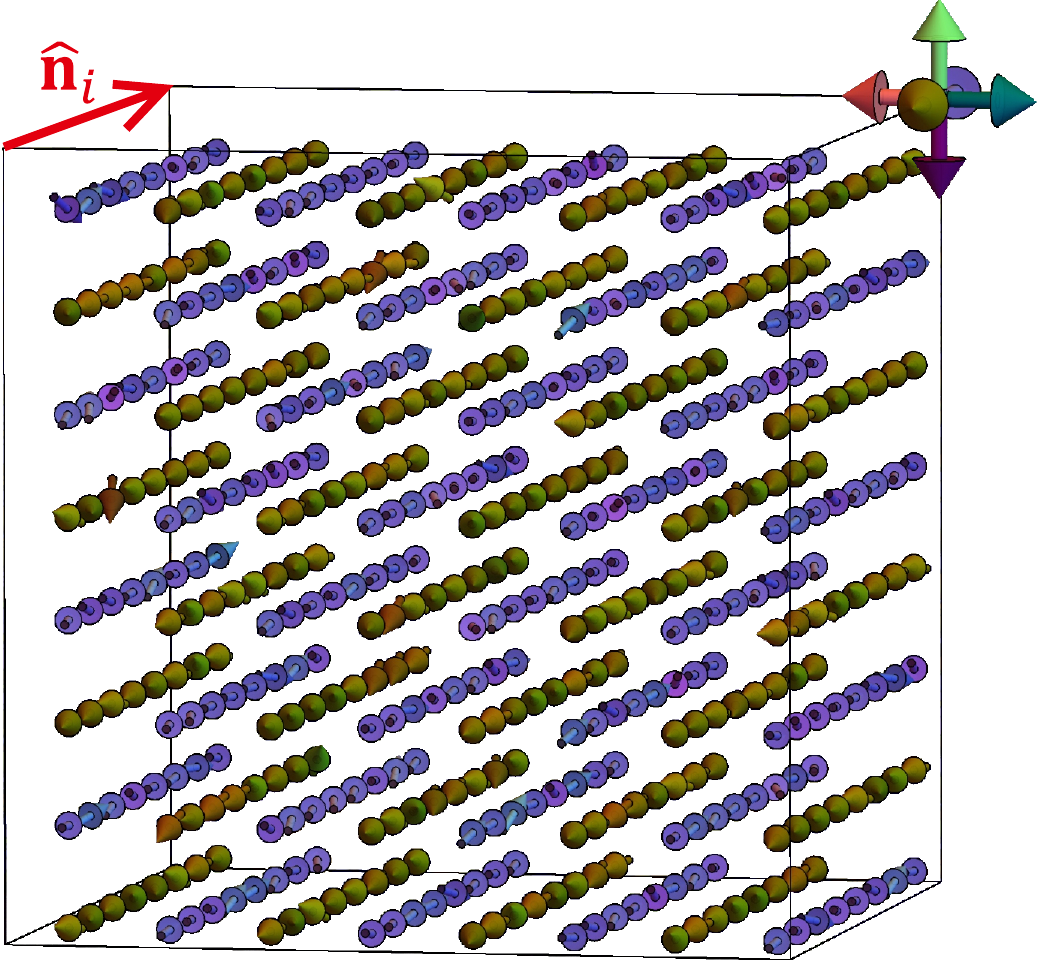}\\(c) $\lambda_e = 1.0$, $\chi_{\|} = 0.16$}
\end{minipage}
\caption{Simulation snapshots of the model system with $\alpha = 0$ and
$\sigma = 20$. Results are shown for parallel configurations with (a) $\lambda_e = 0.3$, (b) $\lambda_e = 0.5$ and (c) $\lambda_e = 1.0$. Different arrow colors correspond to different orientations of magnetic moments. The viewing angle on the system is changed to more clarity.}
\label{fig:chi_snapshots}
\end{figure}

\noindent  The potential reason for this behavior is that the total magnetic moment of the system decreases due to the appearance of magnetically compensated structures. This suggestion can be confirmed from a visual examination of the simulation snapshots given in Fig. \ref{fig:chi_snapshots}. The viewing angle on the system is changed to provide more clarity: the blue vectors correspond to the direction of the $Oz$ axis. All systems have a high value of the anisotropy parameter $\sigma = 20$, which leads to a rigid alignment of the magnetic moments along the easy axes. For the system with $\lambda_e=0.3$ (Fig. \ref{fig:chi_snapshots} (a)), there are no regular structures in the direction of the magnetic moments: the snapshot contains many possible alignment variations of the magnetic moments in inner chains along the easy axis. Fig. \ref{fig:chi_snapshots}~(b) shows that for the system with $\lambda_e=0.5$, a pronounced arrangement of the magnetic moments in chains along the easy axes is present: these are directed mainly in one direction within a single chain. Increasing the intensity of dipole interactions up to $\lambda_e = 1$ (Fig. \ref{fig:chi_snapshots} (c)), we note an additional tendency in the antiparallel alignment of chains along the easy axes, which leads to a further decrease in the system susceptibility. This case  requires some discussion. The plane perpendicular to $\hat{\bm{n}_i}$ in Fig. \ref{fig:chi_snapshots} (c) has a clear checkerboard pattern, since there are four antiparallel neighbors for each chain parallel to $\hat{\bm{n}_i}$. For a single  nanoparticle, it is possible to conclude the following about its six nearest neighbors: two form the most conducive orientation ``head-to-tail''; the next four have the second-most advantageous orientation ``side-by-side''. The magnetic response of such a system is very weak and needs a strong applied magnetic field to obtain the magneto-active material.

\begin{figure}[t!]
\center
\includegraphics[width=\linewidth]{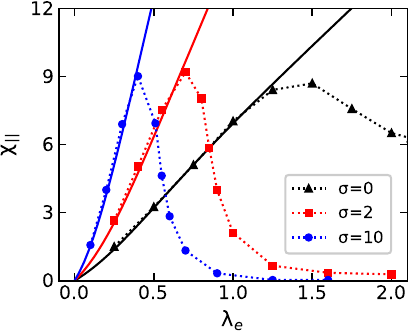}
\caption{The initial magnetic susceptibility $\chi_{||}$ as a function of parameter $\lambda_e$ for parallel configuration. The solid lines correspond to the new theory for $\chi_{||}$. The symbols are from the Monte-Carlo simulations. The results are shown for systems with $\sigma = 0$ (black triangles and lines), $\sigma = 2$ (red squares and lines) and $\sigma = 10$ (blue circles and lines).}
\label{fig:chi_lambda}
\end{figure}

From Fig. \ref{fig:chi_snapshots}, one can conclude that the dependency of $\chi_{\|}$ as a function of $\lambda_e$ at a fixed value of $\sigma$ is also non-monotonic. This fact was observed in Ref. \cite{Solovyova2020} at $\sigma=0$: the Molecular Dynamic simulation results are denoted in Fig. \ref{fig:chi_lambda} by the black symbols with the dashed black curve. The excellent agreement between the new theory (\ref{chi}) and the simulation data up to $\lambda_e \sim 1.25$ should be noted, when there are no structural transformations in the magnetic systems. A further increase in interaction intensity leads to a decrease in susceptibility, the reason for which is the particular behavior of the dipole ordering: almost all the magnetic moments are aligned in long antiparallel chains along the direction of the easy axis. During the current investigation, the dependency of $\chi_{\|}$ as a function of $\lambda_e$ was calculated for systems with an easy axis, the anisotropy parameters of which had the values $\sigma = 2$ and 10. The obtained MC results are denoted in Fig. \ref{fig:chi_lambda} as red symbols with a dashed red curve and blue symbols with a dashed blue curve, respectively. The increasing of $\sigma$ causes a shift in the susceptibility maximum to the left. The value of this shift from $\sigma = 0$ to 2 is greater than from $\sigma = 2$ to 10. It should also be stressed that susceptibility begins to decline when the value $\chi _{\|} \simeq 9$ is reached, regardless of the value of the anisotropy parameter $\sigma$. This is consistent with the results from Fig. \ref{fig:chi_mc} (c): susceptibility increases as $\sigma$ increases up to the value of $\chi_{\|} \simeq 9$ and then decreases further. This fact allows us to conclude that the model system can demonstrate the maximum value of susceptibility in the parallel configuration $\chi _{\|} \simeq 9$, and that this is an inherent property of the considered nanoparticle placement on the nodes of the simple cubic lattice. It should be emphasized that in a system of interacting immobilized superparamagnetic  nanoparticles located randomly, the pronounced regular structuring of the magnetic moments is not observed with the parameters under consideration \cite{Elfimova2019}.

\begin{figure}[t!]
\center
\includegraphics[width=\linewidth]{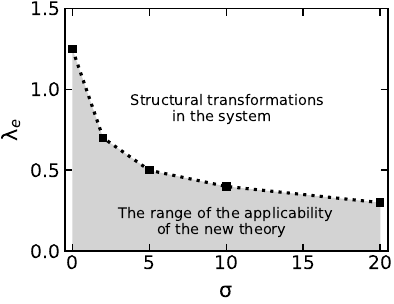}
\caption{The phase diagram of the uniaxial superparamagnetic nanoparticles embedded in a simple cubic lattice. The grey area corresponds to the state of a system without the regular structuring of the particle magnetic moments, where the new theory (\ref{chi}), (\ref{dF_log_par}), (\ref{dF_log_perp}) is able to describe the magnetic properties of the model.  }
\label{fig:phase_diag}
\end{figure}

The new theory is able to predict the magnetic properties of the model system over a range of parameters $\lambda_e$ and $\sigma$, for which the susceptibility value is  $\chi_{\|} \leq  9$ and there is no regular structuring of nanoparticle magnetic moments. To clarify this range, it is possible to plot a phase diagram using the MC data from Fig. \ref{fig:chi_mc}:
\begin{itemize}
    \item at $\lambda_e=0.3$ new theory works well up to $\sigma=20$;
    \item at $\lambda_e=0.5$ new theory works well up to $\sigma=5$;
\end{itemize}
\noindent and from Fig. \ref{fig:chi_lambda}:
\begin{itemize}
    \item at $\sigma=0$ new theory works well up to $\lambda_e=1.25$;
    \item at $\sigma=2$ new theory works well up to $\lambda_e=0.7$;
    \item at $\sigma=10$ new theory works well up to $\lambda_e=0.4$.
\end{itemize}
\noindent The range of the applicability of the new theory is denoted as the grey area in Fig. \ref{fig:phase_diag}. The area above the dashed line indicates an ordered state with the presence of ferromagnetic chains arranged antiferromagnetically in zero fields, which leads to a decrease in susceptibility. The antiferromagnetic ordering of the magnetic moments was also discovered in a study of the ground state of dipoles embedded in simple cubic lattice nodes in the absence of an external magnetic field \cite{kretschmer1979ordering}.

The deviation of new theory from the results of computer modeling in Figs. \ref{fig:chi_mc} (c), \ref{fig:chi_lambda} is connected with the consideration of only the pair correlations in the system and the truncation of the Helmholtz free energy expansions (\ref{dF_log_par}), (\ref{dF_log_perp}) up to the third power of the effective dipolar coupling constant. To expand the given scope of the new theory, one can try to increase the number of terms in the virial series. However, when applying the virial expansion method, it is difficult to construct a theoretical approach valid for systems with a strong dipolar regime. In this case, an alternative method can be used, for instance \cite{Budkov2018, Budkov2016}.

\section{Conclusion}\label{sec:concl}

The static magnetic response of a simple cubic lattice of interacting superparamagnetic nanoparticles has been studied via theory and simulation. The potential energy of the system includes one-particle dipole-easy axis interaction, one-particle dipole-field interaction, and long-range interparticle dipole-dipole interactions. Two orientational distributions of the easy axis in the system have been considered: aligned parallel and perpendicular to the direction of an external magnetic field. For both cases, a theoretical expression for the Helmholtz free energy in logarithmic form has been derived considering pairwise dipole-dipole interactions from the rigorous methods of statistical physics. Using the obtained formula for the Helmholtz free energy, the magnetic properties have been  investigated over a broad range of parameters. The theoretical predictions have been critically compared with MC simulation data.

For the parallel configuration, the magnetic moments are preferably aligned along the direction of the field, which strongly enhances magnetization. For the perpendicular configuration, the magnetic moments are held by the easy axes perpendicular to the field: magnetization decreases as the anisotropy parameter $\sigma$ increases. In both cases, interparticle dipole-dipole interactions lead to an increase in magnetization, but this enhancement is much greater for the parallel  configuration. The ideal approach is not capable of describing the simulation data adequately, whereas the theory examined in this paper (which takes into account interparticle interactions) is much more efficient.

The susceptibility curves from the MC simulations have demonstrated several  interesting new results. Firstly, non-monotonic behavior of $\chi_{\|}$ has been observed as the anisotropy parameter or the dipole-dipole interaction intensity increase. For a  system of  interacting  immobilized  superparamagnetic   nanoparticles  located randomly, the initial magnetic susceptibility increases monotonically as the anisotropy parameter or the dipole-dipole interaction intensity increase \cite{Elfimova2019}. Secondly, for any set of intrinsic parameters, the maximum achievable value of $\chi_{\|}$ is equal to $\simeq 9$. This allows us to conclude that the  system under consideration undergoes a phase transition in the ordering of magnetic moments. The phase diagram of the orientational state of the magnetic moments was plotted along the $\lambda_e$ -- $\sigma$ axes, which shows the region with regular orientational structuring of the magnetic moments and the region lacking any orientational structures. The theoretic susceptibility curves are in strong agreement with the MC data only for parameters that correspond to the system without regular structuring of the magnetic moments. The obtained results are important for the development of functional magnetic materials with controlled properties.

\section*{Acknowledgements}
The reported study was funded by RFBR, project number 20-02-00358.

\section*{Appendix A: Parallel Configuration}
\setcounter{equation}{0}
\renewcommand{\theequation}{A.\arabic{equation}}

For the parallel configuration, the results of averaging the coefficients $b_1$, $b_2$, $b_3$ over the magnetic moment orientations can be written as follows:

\begin{eqnarray}
b_1 &=&  2 Q_1^2(\alpha,\sigma) \gamma_{12}, \\
b_2 &=&  \frac{36}{35} Q_2^2(\alpha,\sigma)\gamma_{24}  \nonumber \\
&+& \frac{2}{3} Q_2(\alpha,\sigma) \left( 1-\frac{Q_2(\alpha,\sigma)}{7} \right) \gamma_{22}  \nonumber \\
&+& \frac{1}{3} \left(1+ \frac{Q_2^2(\alpha,\sigma)}{5} \right) \gamma_{20},\\
b_3 &=&  \frac{6}{77} \big(3 Q_1(\alpha,\sigma) - 5 Q_3 (\alpha,\sigma) \big)^2\gamma_{36}  \nonumber \\
 &+& \frac{18}{11} \bigg( \frac{4Q_1(\alpha,\sigma)Q_3(\alpha,\sigma)}{5}  - \frac{3 Q_1^2(\alpha,\sigma)}{7} \nonumber \\
 &-& \frac{Q_3^2(\alpha,\sigma)}{7}\bigg) \gamma_{34}+ \frac{1}{7} \left(3 Q_1^2(\alpha,\sigma) + Q_3^2(\alpha,\sigma) \right) \gamma_{32}  \nonumber \\
 &+& \frac{1}{42} \bigg(\frac{6 Q_1(\alpha,\sigma)Q_3(\alpha,\sigma)}{5} + 3 Q_1^2(\alpha,\sigma) \nonumber \\
 &-& Q_3^2(\alpha,\sigma)\bigg) \gamma_{30} ,
\end{eqnarray}
\noindent where several functions have been introduced as
\begin{eqnarray}
Q_1(\alpha,\sigma) &=& \frac{1}{2 Q_0(\alpha, \sigma)} \int \limits_{-1}^1 \exp \left(\alpha t  + \sigma t ^2  \right) t \ d t,  \ \ \ \\
Q_1(\alpha,0) &=& L(\alpha), \nonumber \\
L(\alpha) &=& \coth \alpha - \frac{1}{\alpha},\nonumber \\
Q_2(\alpha,\sigma) &=&  \frac{3}{4 Q_0(\alpha, \sigma)} \int \limits_{-1}^1 \exp \left(\alpha t  + \sigma t ^2  \right) t^2 \ d t
\nonumber \\
&-&\frac{1}{2}, \ \ \ \ \ \ \ \\
Q_2(\alpha,0) &=& L_3(\alpha),\nonumber \\
L_3(\alpha) &=& 1-3\frac{L(\alpha)}{\alpha},\nonumber \\
Q_3(\alpha,\sigma) &=& \frac{1}{2 Q_0(\alpha, \sigma)} \int \limits_{-1}^1 \exp \left(\alpha t  + \sigma t ^2  \right) t^3 \ d t,  \ \ \  \ \ \  \\
Q_3(\alpha,0) &=&L(\alpha) - \frac{2}{\alpha} L_3(\alpha). \nonumber
\end{eqnarray}
The numbers $\gamma_{pq}$ involve the sum of position-dependent expressions over the nodes of the cubic lattice, limited by cylinder size:
\begin{eqnarray}\label{gamma}
\gamma_{pq} &=& \sum \limits _  {j=2} ^N \frac{1}{\tilde{r}_{1j}^{3p}} P_q \left( \frac{\tilde{z}_{1j}}{\tilde{r}_{1j}} \right),
 \end{eqnarray}
\noindent This definition introduces the dimensionless interparticle separation vector $\tilde{\textbf{r}}_{1j} = \textbf{r}_{1j} / a $, where $\tilde{z}_{1j}$ is the \emph{z}-component of vector $\tilde{\textbf{r}}_{1j}$ in the laboratory coordinate system shown in Fig. \ref{fig:config} (a). $P_q$ $(q=0,2,4,6)$ in (\ref{gamma}) denote Legendre polynomials.
It is assumed that the  nanoparticle \emph{1} is fixed at the origin of the laboratory frame. All the other nodes of the simple cubic lattice except $(\tilde{x}_{1}, \tilde{y}_{1},\tilde{z}_{1}) \equiv (0,0,0)$, can be occupied by nanoparticle~\emph{j}
\begin{eqnarray}
-R \leqslant & \tilde{x}_{j}\leqslant& R,\nonumber  \\
-R \leqslant& \tilde{y}_{j}\leqslant & R, \nonumber \\
-h R \leqslant& \tilde{z}_{j}\leqslant & h R, \\
\left(\tilde{x}_{j}\right)^2 + & \left(\tilde{y}_{j}\right)^2 & \leqslant R^2,\nonumber\\
\left(\tilde{x}_{j}\right)^2 + & \left(\tilde{y}_{j}\right)^2&+\left(\tilde{z}_{j}\right)^2>0.\nonumber
\end{eqnarray}
\noindent The cylinder is limited by the dimensionless radius \emph{R} and the height $2h$. The results for $\gamma_{pq}$ obtained for model system with $N \simeq 25\times 10^6$ ferroparticles are $\gamma_{12} = 2.0944$, $\gamma_{24}=3.2257$, $\gamma_{22}=\gamma_{32}=0$, $\gamma_{20}=8.4016$, $\gamma_{36}=0.6553$, $\gamma_{34}=3.4081$, $\gamma_{30}=6.6289$.

\section*{Appendix B: Perpendicular Configuration}
\setcounter{equation}{0}
\renewcommand{\theequation}{B.\arabic{equation}}
For the perpendicular configuration, the results of averaging the coefficients $b_1$, $b_2$, $b_3$ over magnetic moment orientations can be written as follows:
\begin{eqnarray}
b_1 &=&  2 R_1^2(\alpha,\sigma) \epsilon_{1}, \\
b_2 &=& \left(R_2^2(\alpha,\sigma) + 3 R_3^2(\alpha,\sigma) - 2 R_3(\alpha,\sigma) + 1 \right)\epsilon_{2} \nonumber \\
&+& \frac{9}{4} \left( 2 R_3(\alpha,\sigma) - 3 R_3^2(\alpha,\sigma) - R_2^2(\alpha,\sigma) + 1\right)\epsilon_{3}\nonumber \\
&+& \frac{9}{4} \epsilon_{3}, \ \ \  \ \ \  \  \\
b_3 &=& \frac{4}{3} R_5^2(\alpha,\sigma) \epsilon_{4} \nonumber \\
&+&  4 \left[ (R_4(\alpha,\sigma) - R_5(\alpha,\sigma))^2 + R_6^2(\alpha,\sigma) \right] \epsilon_{5} \nonumber \\
&+& 18 \big[ R_4^2(\alpha,\sigma)+ R_6^2(\alpha,\sigma) + R_5(\alpha,\sigma)R_6(\alpha,\sigma) \nonumber \\
&-& R_4(\alpha,\sigma)R_5(\alpha,\sigma)\big] \epsilon_{6} \nonumber \\
&+& 9 (R_4(\alpha,\sigma) - R_5(\alpha,\sigma)) R_6(\alpha,\sigma) \epsilon_{7},
\end{eqnarray}

\noindent where additional functions have been introduced as

\begin{eqnarray}
R_1(\alpha,\sigma) &=&  \int \limits_{0}^1 \exp \left(\sigma t ^2  \right) I_0 (\alpha \sqrt{1-t^2}) \sqrt{1-t^2} d t\nonumber \\
&\times& \left[R_0(\alpha,\sigma)\right]^{-1} ,  \\
 R_1(\alpha,0) &=&L(\alpha), \nonumber \\
R_2(\alpha,\sigma) &=& \int \limits_{0}^1 \exp \left(\sigma t ^2  \right) I_2 (\alpha \sqrt{1-t^2}) \left(1-t^2\right) d t \nonumber \\
&\times& \left[R_0(\alpha,\sigma)\right]^{-1},  \\
 R_2(\alpha,0) &=&  L_3(\alpha) , \nonumber\\
R_3(\alpha,\sigma) &=& \int \limits_{0}^1 \exp \left(\sigma t ^2  \right) I_0 (\alpha \sqrt{1-t^2}) t^2 d t \nonumber\\
&\times& \left[R_0(\alpha,\sigma)\right]^{-1} ,
 \end{eqnarray}
\begin{eqnarray}
R_3(\alpha,0) &=& \frac{L(\alpha)}{\alpha} , \nonumber \\
R_4(\alpha,\sigma) &=&  \int \limits_{0}^1 \exp \left(\sigma t ^2  \right) I_1 (\alpha \sqrt{1-t^2}) \sqrt{1-t^2}^{3} d t \nonumber\\
&\times& \left[R_0(\alpha,\sigma)\right]^{-1} ,  \\
R_4(\alpha,0) &=& L(\alpha) - \frac{L_3(\alpha)}{\alpha}, \nonumber \\
R_5(\alpha,\sigma) &=&  \int \limits_{0}^1 \exp \left(\sigma t ^2  \right) I_3 (\alpha \sqrt{1-t^2}) \sqrt{1-t^2}^{3} d t\nonumber\\
&\times& \left[R_0(\alpha,\sigma)\right]^{-1} ,  \\
R_5(\alpha,0) &=&  L(\alpha) - 2\frac{L_3(\alpha)}{\alpha}, \nonumber\\
R_6(\alpha,\sigma) &=&  \int \limits_{0}^1 \exp \left(\sigma t ^2  \right) I_1 (\alpha \sqrt{1-t^2}) \sqrt{1-t^2} t^2 d t\nonumber\\
&\times& \left[R_0(\alpha,\sigma)\right]^{-1} ,  \\
R_6(\alpha,0) &=&  \frac{L_3(\alpha)}{\alpha}. \nonumber
\end{eqnarray}
Here, $I_k(x)$ is the modified Bessel function of \emph{k} order, the numbers $\epsilon_k$ involve the sum of position-dependent expressions over the nodes of the cubic lattice, limited by cylinder size:
\begin{eqnarray}
\epsilon_{1} &=& \sum \limits _  {j=2} ^N \frac{1}{\tilde{r}_{1j}^{3}} P_2 \left( \frac{\tilde{x}_{1j}}{\tilde{r}_{1j}} \right) \simeq 2.0944, \\
\epsilon_{2} &=& \sum \limits _  {j=2} ^N \frac{1}{\tilde{r}_{1j}^{6}} P_2^2 \left( \frac{\tilde{x}_{1j}}{\tilde{r}_{1j}} \right) \simeq 3.3393, \\
\epsilon_{3} &=& \sum \limits _  {j=2} ^N \frac{\tilde{x}_{1j}^2\tilde{z}_{1j}^2}{\tilde{r}_{1j}^{10}} \simeq 0.1915, \\
\epsilon_{4} &=& \sum \limits _  {j=2} ^N \frac{1}{\tilde{r}_{1j}^{9}} P_2^3 \left( \frac{\tilde{x}_{1j}}{\tilde{r}_{1j}} \right) \simeq 1.4880,\\
\epsilon_{5} &=& \sum \limits _  {j=2} ^N \frac{1}{\tilde{r}_{1j}^{9}} P_2 \left( \frac{\tilde{x}_{1j}}{\tilde{r}_{1j}}  \right) P_2^2 \left( \frac{\tilde{z}_{1j}}{\tilde{r}_{1j}} \right)\nonumber \\
&\simeq& -0.7440, \\
\epsilon_{6} &=& \sum \limits _  {j=2} ^N\frac{\tilde{x}_{1j}^2\tilde{z}_{1j}^2}{\tilde{r}_{1j}^{13}} P_2 \left( \frac{\tilde{z}_{1j}}{\tilde{r}_{1j}} \right) \simeq 0.0114, \\
\epsilon_{7} &=& \sum \limits _  {j=2} ^N\frac{\tilde{y}_{1j}^2\tilde{z}_{1j}^2 (8\tilde{x}_{1j}^2 - \tilde{y}_{1j}^2 - \tilde{z}_{1j}^2)}{\tilde{r}_{1j}^{15}} \nonumber \\
&\simeq& -0.0318. \ \ \ \
\end{eqnarray}

%\section*{References}

\bibliography{ref}

\begin{thebibliography}{48}%
\makeatletter
\providecommand \@ifxundefined [1]{%
 \@ifx{#1\undefined}
}%
\providecommand \@ifnum [1]{%
 \ifnum #1\expandafter \@firstoftwo
 \else \expandafter \@secondoftwo
 \fi
}%
\providecommand \@ifx [1]{%
 \ifx #1\expandafter \@firstoftwo
 \else \expandafter \@secondoftwo
 \fi
}%
\providecommand \natexlab [1]{#1}%
\providecommand \enquote  [1]{``#1''}%
\providecommand \bibnamefont  [1]{#1}%
\providecommand \bibfnamefont [1]{#1}%
\providecommand \citenamefont [1]{#1}%
\providecommand \href@noop [0]{\@secondoftwo}%
\providecommand \href [0]{\begingroup \@sanitize@url \@href}%
\providecommand \@href[1]{\@@startlink{#1}\@@href}%
\providecommand \@@href[1]{\endgroup#1\@@endlink}%
\providecommand \@sanitize@url [0]{\catcode `\\12\catcode `\$12\catcode
  `\&12\catcode `\#12\catcode `\^12\catcode `\_12\catcode `\%12\relax}%
\providecommand \@@startlink[1]{}%
\providecommand \@@endlink[0]{}%
\providecommand \url  [0]{\begingroup\@sanitize@url \@url }%
\providecommand \@url [1]{\endgroup\@href {#1}{\urlprefix }}%
\providecommand \urlprefix  [0]{URL }%
\providecommand \Eprint [0]{\href }%
\providecommand \doibase [0]{http://dx.doi.org/}%
\providecommand \selectlanguage [0]{\@gobble}%
\providecommand \bibinfo  [0]{\@secondoftwo}%
\providecommand \bibfield  [0]{\@secondoftwo}%
\providecommand \translation [1]{[#1]}%
\providecommand \BibitemOpen [0]{}%
\providecommand \bibitemStop [0]{}%
\providecommand \bibitemNoStop [0]{.\EOS\space}%
\providecommand \EOS [0]{\spacefactor3000\relax}%
\providecommand \BibitemShut  [1]{\csname bibitem#1\endcsname}%
\let\auto@bib@innerbib\@empty
%</preamble>
\bibitem [{\citenamefont {Kuznetsova}\ \emph {et~al.}(2019)\citenamefont
  {Kuznetsova}, \citenamefont {Kolesov}, \citenamefont {Fionov}, \citenamefont
  {Kramarenko}, \citenamefont {Stepanov}, \citenamefont {Mikheev},
  \citenamefont {Verona},\ and\ \citenamefont {Solodov}}]{Kuznetsova2019}%
  \BibitemOpen
  \bibfield  {author} {\bibinfo {author} {\bibfnamefont {I.E.}\ \bibnamefont
  {Kuznetsova}}, \bibinfo {author} {\bibfnamefont {V.V.}\ \bibnamefont
  {Kolesov}}, \bibinfo {author} {\bibfnamefont {A.S.}\ \bibnamefont {Fionov}},
  \bibinfo {author} {\bibfnamefont {E.Y.}\ \bibnamefont {Kramarenko}}, \bibinfo
  {author} {\bibfnamefont {G.V.}\ \bibnamefont {Stepanov}}, \bibinfo {author}
  {\bibfnamefont {M.G.}\ \bibnamefont {Mikheev}}, \bibinfo {author}
  {\bibfnamefont {E.}~\bibnamefont {Verona}}, \ and\ \bibinfo {author}
  {\bibfnamefont {I.}~\bibnamefont {Solodov}},\ }\bibfield  {title} {\enquote
  {\bibinfo {title} {Magnetoactive elastomers with controllable radio-absorbing
  properties},}\ }\href {\doibase 10.1016/j.mtcomm.2019.100610} {\bibfield
  {journal} {\bibinfo  {journal} {Materials Today Communications}\ }\textbf
  {\bibinfo {volume} {21}},\ \bibinfo {pages} {100610} (\bibinfo {year}
  {2019})}\BibitemShut {NoStop}%
\bibitem [{\citenamefont {Borin}\ \emph {et~al.}(2019)\citenamefont {Borin},
  \citenamefont {Stepanov},\ and\ \citenamefont {Dohmen}}]{Borin2019105}%
  \BibitemOpen
  \bibfield  {author} {\bibinfo {author} {\bibfnamefont {D.}~\bibnamefont
  {Borin}}, \bibinfo {author} {\bibfnamefont {G.}~\bibnamefont {Stepanov}}, \
  and\ \bibinfo {author} {\bibfnamefont {E.}~\bibnamefont {Dohmen}},\
  }\bibfield  {title} {\enquote {\bibinfo {title} {Hybrid magnetoactive
  elastomer with a soft matrix and mixed powder},}\ }\href {\doibase
  10.1007/s00419-018-1456-9} {\bibfield  {journal} {\bibinfo  {journal}
  {Archive of Applied Mechanics}\ }\textbf {\bibinfo {volume} {89}},\ \bibinfo
  {pages} {105--117} (\bibinfo {year} {2019})}\BibitemShut {NoStop}%
\bibitem [{\citenamefont {Filipcsei}\ \emph {et~al.}(2007)\citenamefont
  {Filipcsei}, \citenamefont {Csetneki}, \citenamefont {Szil{\'a}gyi},\ and\
  \citenamefont {Zr{\'i}nyi}}]{Filipcsei2007}%
  \BibitemOpen
  \bibfield  {author} {\bibinfo {author} {\bibfnamefont {G.}~\bibnamefont
  {Filipcsei}}, \bibinfo {author} {\bibfnamefont {I.}~\bibnamefont {Csetneki}},
  \bibinfo {author} {\bibfnamefont {A.}~\bibnamefont {Szil{\'a}gyi}}, \ and\
  \bibinfo {author} {\bibfnamefont {M.}~\bibnamefont {Zr{\'i}nyi}},\ }\bibfield
   {title} {\enquote {\bibinfo {title} {Magnetic field-responsive smart polymer
  composites},}\ }\href {\doibase 10.1007/12_2006_104} {\bibfield  {journal}
  {\bibinfo  {journal} {Advances in Polymer Science}\ }\textbf {\bibinfo
  {volume} {206}},\ \bibinfo {pages} {137--189} (\bibinfo {year}
  {2007})}\BibitemShut {NoStop}%
\bibitem [{\citenamefont {Blyakhman}\ \emph {et~al.}(2019)\citenamefont
  {Blyakhman}, \citenamefont {Makarova}, \citenamefont {Fadeyev}, \citenamefont
  {Lugovets}, \citenamefont {Safronov}, \citenamefont {Shabadrov},
  \citenamefont {Shklyar}, \citenamefont {Melnikov}, \citenamefont {Orue},\
  and\ \citenamefont {Kurlyandskaya}}]{Blyakhman2019}%
  \BibitemOpen
  \bibfield  {author} {\bibinfo {author} {\bibfnamefont {F.A.}\ \bibnamefont
  {Blyakhman}}, \bibinfo {author} {\bibfnamefont {E.B.}\ \bibnamefont
  {Makarova}}, \bibinfo {author} {\bibfnamefont {F.A.}\ \bibnamefont
  {Fadeyev}}, \bibinfo {author} {\bibfnamefont {D.V.}\ \bibnamefont
  {Lugovets}}, \bibinfo {author} {\bibfnamefont {A.P.}\ \bibnamefont
  {Safronov}}, \bibinfo {author} {\bibfnamefont {P.A.}\ \bibnamefont
  {Shabadrov}}, \bibinfo {author} {\bibfnamefont {T.F.}\ \bibnamefont
  {Shklyar}}, \bibinfo {author} {\bibfnamefont {G.Y.}\ \bibnamefont
  {Melnikov}}, \bibinfo {author} {\bibfnamefont {I.}~\bibnamefont {Orue}}, \
  and\ \bibinfo {author} {\bibfnamefont {G.V.}\ \bibnamefont {Kurlyandskaya}},\
  }\bibfield  {title} {\enquote {\bibinfo {title} {The contribution of magnetic
  nanoparticles to ferrogel biophysical properties},}\ }\href {\doibase
  10.3390/nano9020232} {\bibfield  {journal} {\bibinfo  {journal}
  {Nanomaterials}\ }\textbf {\bibinfo {volume} {9}},\ \bibinfo {pages} {232}
  (\bibinfo {year} {2019})}\BibitemShut {NoStop}%
\bibitem [{\citenamefont {Dutz}\ and\ \citenamefont {Hergt}(2013)}]{Dutz2013}%
  \BibitemOpen
  \bibfield  {author} {\bibinfo {author} {\bibfnamefont {S.}~\bibnamefont
  {Dutz}}\ and\ \bibinfo {author} {\bibfnamefont {R.}~\bibnamefont {Hergt}},\
  }\bibfield  {title} {\enquote {\bibinfo {title} {Magnetic nanoparticle
  heating and heat transfer on a microscale: Basic principles, realities and
  physical limitations of hyperthermia for tumour therapy},}\ }\href {\doibase
  10.3109/02656736.2013.822993} {\bibfield  {journal} {\bibinfo  {journal}
  {International Journal of Hyperthermia}\ }\textbf {\bibinfo {volume} {29}},\
  \bibinfo {pages} {790--800} (\bibinfo {year} {2013})}\BibitemShut {NoStop}%
\bibitem [{\citenamefont {Ortega}\ and\ \citenamefont
  {Pankhurst}(2013)}]{Ortega2013}%
  \BibitemOpen
  \bibfield  {author} {\bibinfo {author} {\bibfnamefont {D.}~\bibnamefont
  {Ortega}}\ and\ \bibinfo {author} {\bibfnamefont {Q.~A.}\ \bibnamefont
  {Pankhurst}},\ }\bibfield  {title} {\enquote {\bibinfo {title} {Magnetic
  hyperthermia},}\ }\href {\doibase 10.1039/9781849734844-00060} {\bibfield
  {journal} {\bibinfo  {journal} {Nanoscience}\ }\textbf {\bibinfo {volume}
  {1}},\ \bibinfo {pages} {60--88} (\bibinfo {year} {2013})}\BibitemShut
  {NoStop}%
\bibitem [{\citenamefont {Zubarev}(2018)}]{Zubarev2018}%
  \BibitemOpen
  \bibfield  {author} {\bibinfo {author} {\bibfnamefont {A.Yu.}\ \bibnamefont
  {Zubarev}},\ }\bibfield  {title} {\enquote {\bibinfo {title} {Magnetic
  hyperthermia in a system of ferromagnetic particles, frozen in a carrier
  medium: Effect of interparticle interactions},}\ }\href {\doibase
  10.1103/PhysRevE.98.032610} {\bibfield  {journal} {\bibinfo  {journal}
  {Physical Review E}\ }\textbf {\bibinfo {volume} {98}},\ \bibinfo {pages}
  {032610} (\bibinfo {year} {2018})}\BibitemShut {NoStop}%
\bibitem [{\citenamefont {Zubarev}(2019)}]{Zubarev2019}%
  \BibitemOpen
  \bibfield  {author} {\bibinfo {author} {\bibfnamefont {A.Yu.}\ \bibnamefont
  {Zubarev}},\ }\bibfield  {title} {\enquote {\bibinfo {title} {Magnetic
  hyperthermia in a system of immobilized magnetically interacting
  particles},}\ }\href {\doibase 10.1103/PhysRevE.99.062609} {\bibfield
  {journal} {\bibinfo  {journal} {Physical Review E}\ }\textbf {\bibinfo
  {volume} {99}},\ \bibinfo {pages} {062609} (\bibinfo {year}
  {2019})}\BibitemShut {NoStop}%
\bibitem [{\citenamefont {Zhang}\ \emph {et~al.}(2017)\citenamefont {Zhang},
  \citenamefont {Zuo}, \citenamefont {Niu}, \citenamefont {Wu}, \citenamefont
  {Wang}, \citenamefont {Guan},\ and\ \citenamefont {Silva}}]{Zhang201713929}%
  \BibitemOpen
  \bibfield  {author} {\bibinfo {author} {\bibfnamefont {W.}~\bibnamefont
  {Zhang}}, \bibinfo {author} {\bibfnamefont {X.}~\bibnamefont {Zuo}}, \bibinfo
  {author} {\bibfnamefont {Y.}~\bibnamefont {Niu}}, \bibinfo {author}
  {\bibfnamefont {C.}~\bibnamefont {Wu}}, \bibinfo {author} {\bibfnamefont
  {S.}~\bibnamefont {Wang}}, \bibinfo {author} {\bibfnamefont {S.}~\bibnamefont
  {Guan}}, \ and\ \bibinfo {author} {\bibfnamefont {S.R.P.}\ \bibnamefont
  {Silva}},\ }\bibfield  {title} {\enquote {\bibinfo {title} {Novel
  nanoparticles with cr3+ substituted ferrite for self-regulating temperature
  hyperthermia},}\ }\href {\doibase 10.1039/c7nr02356a} {\bibfield  {journal}
  {\bibinfo  {journal} {Nanoscale}\ }\textbf {\bibinfo {volume} {9}},\ \bibinfo
  {pages} {13929--13937} (\bibinfo {year} {2017})}\BibitemShut {NoStop}%
\bibitem [{\citenamefont {Piehler}\ \emph {et~al.}(2020)\citenamefont
  {Piehler}, \citenamefont {D\"{a}hring}, \citenamefont {Grandke},
  \citenamefont {G\"{o}ring}, \citenamefont {Couleaud}, \citenamefont {Aires},
  \citenamefont {Cortajarena}, \citenamefont {Courty}, \citenamefont {Latorre},
  \citenamefont {Somoza}, \citenamefont {Teichgr\"{a}ber},\ and\ \citenamefont
  {Hilger}}]{Piehler2020}%
  \BibitemOpen
  \bibfield  {author} {\bibinfo {author} {\bibfnamefont {S.}~\bibnamefont
  {Piehler}}, \bibinfo {author} {\bibfnamefont {H.}~\bibnamefont
  {D\"{a}hring}}, \bibinfo {author} {\bibfnamefont {J.}~\bibnamefont
  {Grandke}}, \bibinfo {author} {\bibfnamefont {J.}~\bibnamefont {G\"{o}ring}},
  \bibinfo {author} {\bibfnamefont {P.}~\bibnamefont {Couleaud}}, \bibinfo
  {author} {\bibfnamefont {A.}~\bibnamefont {Aires}}, \bibinfo {author}
  {\bibfnamefont {A.L.}\ \bibnamefont {Cortajarena}}, \bibinfo {author}
  {\bibfnamefont {J.}~\bibnamefont {Courty}}, \bibinfo {author} {\bibfnamefont
  {A.}~\bibnamefont {Latorre}}, \bibinfo {author} {\bibfnamefont
  {A.}~\bibnamefont {Somoza}}, \bibinfo {author} {\bibfnamefont
  {U.}~\bibnamefont {Teichgr\"{a}ber}}, \ and\ \bibinfo {author} {\bibfnamefont
  {I.}~\bibnamefont {Hilger}},\ }\bibfield  {title} {\enquote {\bibinfo {title}
  {Iron oxide nanoparticles as carriers for dox and magnetic hyperthermia after
  intratumoral application into breast cancer in mice: impact and future
  perspectives},}\ }\href {\doibase 10.3390/nano10061016} {\bibfield  {journal}
  {\bibinfo  {journal} {Nanomaterials}\ }\textbf {\bibinfo {volume} {10}},\
  \bibinfo {pages} {1016} (\bibinfo {year} {2020})}\BibitemShut {NoStop}%
\bibitem [{\citenamefont {Brero}\ \emph {et~al.}(2020)\citenamefont {Brero},
  \citenamefont {Albino}, \citenamefont {Antoccia}, \citenamefont {Arosio},
  \citenamefont {Avolio}, \citenamefont {Berardinelli}, \citenamefont
  {Bettega}, \citenamefont {Calzolari}, \citenamefont {Ciocca}, \citenamefont
  {Corti}, \citenamefont {Facoetti}, \citenamefont {Gallo}, \citenamefont
  {Groppi}, \citenamefont {Guerrini}, \citenamefont {Innocenti}, \citenamefont
  {Lenardi}, \citenamefont {Locarno}, \citenamefont {Manenti}, \citenamefont
  {Marchesini}, \citenamefont {Mariani}, \citenamefont {Orsini}, \citenamefont
  {Pignoli}, \citenamefont {Sangregorio}, \citenamefont {Veronese},\ and\
  \citenamefont {Lascialfari}}]{Brero20201}%
  \BibitemOpen
  \bibfield  {author} {\bibinfo {author} {\bibfnamefont {F.}~\bibnamefont
  {Brero}}, \bibinfo {author} {\bibfnamefont {M.}~\bibnamefont {Albino}},
  \bibinfo {author} {\bibfnamefont {A.}~\bibnamefont {Antoccia}}, \bibinfo
  {author} {\bibfnamefont {P.}~\bibnamefont {Arosio}}, \bibinfo {author}
  {\bibfnamefont {M.}~\bibnamefont {Avolio}}, \bibinfo {author} {\bibfnamefont
  {F.}~\bibnamefont {Berardinelli}}, \bibinfo {author} {\bibfnamefont
  {D.}~\bibnamefont {Bettega}}, \bibinfo {author} {\bibfnamefont
  {P.}~\bibnamefont {Calzolari}}, \bibinfo {author} {\bibfnamefont
  {M.}~\bibnamefont {Ciocca}}, \bibinfo {author} {\bibfnamefont
  {M.}~\bibnamefont {Corti}}, \bibinfo {author} {\bibfnamefont
  {A.}~\bibnamefont {Facoetti}}, \bibinfo {author} {\bibfnamefont
  {S.}~\bibnamefont {Gallo}}, \bibinfo {author} {\bibfnamefont
  {F.}~\bibnamefont {Groppi}}, \bibinfo {author} {\bibfnamefont
  {A.}~\bibnamefont {Guerrini}}, \bibinfo {author} {\bibfnamefont
  {C.}~\bibnamefont {Innocenti}}, \bibinfo {author} {\bibfnamefont
  {C.}~\bibnamefont {Lenardi}}, \bibinfo {author} {\bibfnamefont
  {S.}~\bibnamefont {Locarno}}, \bibinfo {author} {\bibfnamefont
  {S.}~\bibnamefont {Manenti}}, \bibinfo {author} {\bibfnamefont
  {R.}~\bibnamefont {Marchesini}}, \bibinfo {author} {\bibfnamefont
  {M.}~\bibnamefont {Mariani}}, \bibinfo {author} {\bibfnamefont
  {F.}~\bibnamefont {Orsini}}, \bibinfo {author} {\bibfnamefont
  {E.}~\bibnamefont {Pignoli}}, \bibinfo {author} {\bibfnamefont
  {C.}~\bibnamefont {Sangregorio}}, \bibinfo {author} {\bibfnamefont
  {I.}~\bibnamefont {Veronese}}, \ and\ \bibinfo {author} {\bibfnamefont
  {A.}~\bibnamefont {Lascialfari}},\ }\bibfield  {title} {\enquote {\bibinfo
  {title} {Hadron therapy, magnetic nanoparticles and hyperthermia: A promising
  combined tool for pancreatic cancer treatment},}\ }\href {\doibase
  10.3390/nano10101919} {\bibfield  {journal} {\bibinfo  {journal}
  {Nanomaterials}\ }\textbf {\bibinfo {volume} {10}},\ \bibinfo {pages} {1919}
  (\bibinfo {year} {2020})}\BibitemShut {NoStop}%
\bibitem [{\citenamefont {Raikher}\ and\ \citenamefont
  {Shliomis}(1974)}]{Raikher1974}%
  \BibitemOpen
  \bibfield  {author} {\bibinfo {author} {\bibfnamefont {Yu.L.}\ \bibnamefont
  {Raikher}}\ and\ \bibinfo {author} {\bibfnamefont {M.I.}\ \bibnamefont
  {Shliomis}},\ }\bibfield  {title} {\enquote {\bibinfo {title} {Theory of
  dispersion of the magnetic susceptibility of fine ferromagnetic particles},}\
  }\href@noop {} {\bibfield  {journal} {\bibinfo  {journal} {Journal of
  Experimental and Theoretical Physics}\ }\textbf {\bibinfo {volume} {40}},\
  \bibinfo {pages} {526–532} (\bibinfo {year} {1974})}\BibitemShut {NoStop}%
\bibitem [{\citenamefont {Shliomis}\ and\ \citenamefont
  {Stepanov}(1994)}]{Shliomis1994}%
  \BibitemOpen
  \bibfield  {author} {\bibinfo {author} {\bibfnamefont {M.I.}\ \bibnamefont
  {Shliomis}}\ and\ \bibinfo {author} {\bibfnamefont {V.I.}\ \bibnamefont
  {Stepanov}},\ }\bibfield  {title} {\enquote {\bibinfo {title} {Theory of the
  dynamic susceptibility of magnetic fluids},}\ }\href {\doibase
  10.1002/9780470141465.ch1} {\bibfield  {journal} {\bibinfo  {journal}
  {Advances in Chemical Physics}\ }\textbf {\bibinfo {volume} {87}},\ \bibinfo
  {pages} {1–30} (\bibinfo {year} {1994})}\BibitemShut {NoStop}%
\bibitem [{\citenamefont {Gervald}\ \emph {et~al.}(2010)\citenamefont
  {Gervald}, \citenamefont {Gritskova},\ and\ \citenamefont
  {Prokopov}}]{Gervald_2010}%
  \BibitemOpen
  \bibfield  {author} {\bibinfo {author} {\bibfnamefont {A.Yu.}\ \bibnamefont
  {Gervald}}, \bibinfo {author} {\bibfnamefont {I.A.}\ \bibnamefont
  {Gritskova}}, \ and\ \bibinfo {author} {\bibfnamefont {N.I.}\ \bibnamefont
  {Prokopov}},\ }\bibfield  {title} {\enquote {\bibinfo {title} {Synthesis of
  magnetic polymeric microspheres},}\ }\href {\doibase
  10.1070/RC2010v079n03ABEH004068} {\bibfield  {journal} {\bibinfo  {journal}
  {Russian Chemical Reviews}\ }\textbf {\bibinfo {volume} {79}},\ \bibinfo
  {pages} {219--229} (\bibinfo {year} {2010})}\BibitemShut {NoStop}%
\bibitem [{\citenamefont {Valiev}\ \emph {et~al.}(2019)\citenamefont {Valiev},
  \citenamefont {Ya~Minaev}, \citenamefont {Stepanov},\ and\ \citenamefont
  {Karnet}}]{Valiev_2019}%
  \BibitemOpen
  \bibfield  {author} {\bibinfo {author} {\bibfnamefont {H.H.}\ \bibnamefont
  {Valiev}}, \bibinfo {author} {\bibfnamefont {A.}~\bibnamefont {Ya~Minaev}},
  \bibinfo {author} {\bibfnamefont {G.V.}\ \bibnamefont {Stepanov}}, \ and\
  \bibinfo {author} {\bibfnamefont {Y.N.}\ \bibnamefont {Karnet}},\ }\bibfield
  {title} {\enquote {\bibinfo {title} {Study of filler microstructure in
  magnetic soft composites},}\ }\href {\doibase
  10.1088/1742-6596/1260/11/112034} {\bibfield  {journal} {\bibinfo  {journal}
  {Journal of Physics: Conference Series}\ }\textbf {\bibinfo {volume}
  {1260}},\ \bibinfo {pages} {112034} (\bibinfo {year} {2019})}\BibitemShut
  {NoStop}%
\bibitem [{\citenamefont {Ganesan}\ \emph {et~al.}(2019)\citenamefont
  {Ganesan}, \citenamefont {Lahiri}, \citenamefont {Louis}, \citenamefont
  {Philip},\ and\ \citenamefont {Damodaran}}]{Ganesan2019315}%
  \BibitemOpen
  \bibfield  {author} {\bibinfo {author} {\bibfnamefont {V.}~\bibnamefont
  {Ganesan}}, \bibinfo {author} {\bibfnamefont {B.B.}\ \bibnamefont {Lahiri}},
  \bibinfo {author} {\bibfnamefont {C.}~\bibnamefont {Louis}}, \bibinfo
  {author} {\bibfnamefont {J.}~\bibnamefont {Philip}}, \ and\ \bibinfo {author}
  {\bibfnamefont {S.P.}\ \bibnamefont {Damodaran}},\ }\bibfield  {title}
  {\enquote {\bibinfo {title} {Size-controlled synthesis of superparamagnetic
  magnetite nanoclusters for heat generation in an alternating magnetic
  field},}\ }\href {\doibase 10.1016/j.molliq.2019.02.095} {\bibfield
  {journal} {\bibinfo  {journal} {Journal of Molecular Liquids}\ }\textbf
  {\bibinfo {volume} {281}},\ \bibinfo {pages} {315--323} (\bibinfo {year}
  {2019})}\BibitemShut {NoStop}%
\bibitem [{\citenamefont {Weeber}\ \emph {et~al.}(2018)\citenamefont {Weeber},
  \citenamefont {Hermes}, \citenamefont {Schmidt},\ and\ \citenamefont
  {Holm}}]{Weeber2018}%
  \BibitemOpen
  \bibfield  {author} {\bibinfo {author} {\bibfnamefont {R.}~\bibnamefont
  {Weeber}}, \bibinfo {author} {\bibfnamefont {M.}~\bibnamefont {Hermes}},
  \bibinfo {author} {\bibfnamefont {A.M.}\ \bibnamefont {Schmidt}}, \ and\
  \bibinfo {author} {\bibfnamefont {C.}~\bibnamefont {Holm}},\ }\bibfield
  {title} {\enquote {\bibinfo {title} {Polymer architecture of magnetic gels: A
  review},}\ }\href {\doibase 10.1088/1361-648X/aaa344} {\bibfield  {journal}
  {\bibinfo  {journal} {Journal of Physics Condensed Matter}\ }\textbf
  {\bibinfo {volume} {30}},\ \bibinfo {pages} {063002} (\bibinfo {year}
  {2018})}\BibitemShut {NoStop}%
\bibitem [{\citenamefont {Tanasa}\ \emph {et~al.}(2019)\citenamefont {Tanasa},
  \citenamefont {Zaharia}, \citenamefont {Radu}, \citenamefont {Surdu},
  \citenamefont {Vasile}, \citenamefont {Damian},\ and\ \citenamefont
  {Andronescu}}]{Tanasa2019}%
  \BibitemOpen
  \bibfield  {author} {\bibinfo {author} {\bibfnamefont {E.}~\bibnamefont
  {Tanasa}}, \bibinfo {author} {\bibfnamefont {C.}~\bibnamefont {Zaharia}},
  \bibinfo {author} {\bibfnamefont {I.-C.}\ \bibnamefont {Radu}}, \bibinfo
  {author} {\bibfnamefont {V.-A.}\ \bibnamefont {Surdu}}, \bibinfo {author}
  {\bibfnamefont {B.S.}\ \bibnamefont {Vasile}}, \bibinfo {author}
  {\bibfnamefont {C.-M.}\ \bibnamefont {Damian}}, \ and\ \bibinfo {author}
  {\bibfnamefont {E.}~\bibnamefont {Andronescu}},\ }\bibfield  {title}
  {\enquote {\bibinfo {title} {Novel nanocomposites based on functionalized
  magnetic nanoparticles and polyacrylamide: Preparation and complex
  characterization},}\ }\href {\doibase 10.3390/nano9101384} {\bibfield
  {journal} {\bibinfo  {journal} {Nanomaterials}\ }\textbf {\bibinfo {volume}
  {9}} (\bibinfo {year} {2019}),\ 10.3390/nano9101384}\BibitemShut {NoStop}%
\bibitem [{\citenamefont {Bastola}\ \emph {et~al.}(2020)\citenamefont
  {Bastola}, \citenamefont {Paudel},\ and\ \citenamefont
  {Li}}]{Bastola2020377}%
  \BibitemOpen
  \bibfield  {author} {\bibinfo {author} {\bibfnamefont {A.K.}\ \bibnamefont
  {Bastola}}, \bibinfo {author} {\bibfnamefont {M.}~\bibnamefont {Paudel}}, \
  and\ \bibinfo {author} {\bibfnamefont {L.}~\bibnamefont {Li}},\ }\bibfield
  {title} {\enquote {\bibinfo {title} {Line-patterned hybrid magnetorheological
  elastomer developed by 3d printing},}\ }\href {\doibase
  10.1177/1045389X19891557} {\bibfield  {journal} {\bibinfo  {journal} {Journal
  of Intelligent Material Systems and Structures}\ }\textbf {\bibinfo {volume}
  {31}},\ \bibinfo {pages} {377--388} (\bibinfo {year} {2020})}\BibitemShut
  {NoStop}%
\bibitem [{\citenamefont {Zakinyan}\ and\ \citenamefont
  {Arefyev}(2020)}]{Zakinyan2020}%
  \BibitemOpen
  \bibfield  {author} {\bibinfo {author} {\bibfnamefont {A.}~\bibnamefont
  {Zakinyan}}\ and\ \bibinfo {author} {\bibfnamefont {I.}~\bibnamefont
  {Arefyev}},\ }\bibfield  {title} {\enquote {\bibinfo {title} {Thermal
  conductivity of emulsion with anisotropic microstructure induced by external
  field},}\ }\href {\doibase 10.1007/s00396-020-04672-x} {\bibfield  {journal}
  {\bibinfo  {journal} {Colloid and Polymer Science}\ }\textbf {\bibinfo
  {volume} {298}},\ \bibinfo {pages} {1063–1076} (\bibinfo {year}
  {2020})}\BibitemShut {NoStop}%
\bibitem [{\citenamefont {Yoshida}\ \emph {et~al.}(2017)\citenamefont
  {Yoshida}, \citenamefont {Matsugi}, \citenamefont {Tsujimura}, \citenamefont
  {Sasayama}, \citenamefont {Enpuku}, \citenamefont {Viereck}, \citenamefont
  {Schilling},\ and\ \citenamefont {Ludwig}}]{Yoshida2017162}%
  \BibitemOpen
  \bibfield  {author} {\bibinfo {author} {\bibfnamefont {T.}~\bibnamefont
  {Yoshida}}, \bibinfo {author} {\bibfnamefont {Y.}~\bibnamefont {Matsugi}},
  \bibinfo {author} {\bibfnamefont {N.}~\bibnamefont {Tsujimura}}, \bibinfo
  {author} {\bibfnamefont {T.}~\bibnamefont {Sasayama}}, \bibinfo {author}
  {\bibfnamefont {K.}~\bibnamefont {Enpuku}}, \bibinfo {author} {\bibfnamefont
  {T.}~\bibnamefont {Viereck}}, \bibinfo {author} {\bibfnamefont
  {M.}~\bibnamefont {Schilling}}, \ and\ \bibinfo {author} {\bibfnamefont
  {F.}~\bibnamefont {Ludwig}},\ }\bibfield  {title} {\enquote {\bibinfo {title}
  {Effect of alignment of easy axes on dynamic magnetization of immobilized
  magnetic nanoparticles},}\ }\href {\doibase 10.1016/j.jmmm.2016.10.040}
  {\bibfield  {journal} {\bibinfo  {journal} {Journal of Magnetism and Magnetic
  Materials}\ }\textbf {\bibinfo {volume} {427}},\ \bibinfo {pages} {162--167}
  (\bibinfo {year} {2017})}\BibitemShut {NoStop}%
\bibitem [{\citenamefont {Elfimova}\ \emph {et~al.}(2019)\citenamefont
  {Elfimova}, \citenamefont {Ivanov},\ and\ \citenamefont
  {Camp}}]{Elfimova2019}%
  \BibitemOpen
  \bibfield  {author} {\bibinfo {author} {\bibfnamefont {E.A.}\ \bibnamefont
  {Elfimova}}, \bibinfo {author} {\bibfnamefont {A.O.}\ \bibnamefont {Ivanov}},
  \ and\ \bibinfo {author} {\bibfnamefont {P.J.}\ \bibnamefont {Camp}},\
  }\bibfield  {title} {\enquote {\bibinfo {title} {Static magnetization of
  immobilized, weakly interacting, superparamagnetic nanoparticles},}\ }\href
  {\doibase 10.1039/C9NR07425B} {\bibfield  {journal} {\bibinfo  {journal}
  {Nanoscale}\ }\textbf {\bibinfo {volume} {11}},\ \bibinfo {pages}
  {21834--21846} (\bibinfo {year} {2019})}\BibitemShut {NoStop}%
\bibitem [{\citenamefont {Socoliuc}\ and\ \citenamefont
  {Popescu}(2020)}]{Socoliuc2020}%
  \BibitemOpen
  \bibfield  {author} {\bibinfo {author} {\bibfnamefont {V.}~\bibnamefont
  {Socoliuc}}\ and\ \bibinfo {author} {\bibfnamefont {L.B.}\ \bibnamefont
  {Popescu}},\ }\bibfield  {title} {\enquote {\bibinfo {title} {Determination
  of the statistics of magnetically induced particle chains in concentrated
  ferrofluids},}\ }\href {\doibase 10.1016/j.jmmm.2020.166532} {\bibfield
  {journal} {\bibinfo  {journal} {Journal of Magnetism and Magnetic Materials}\
  }\textbf {\bibinfo {volume} {502}},\ \bibinfo {pages} {166532} (\bibinfo
  {year} {2020})}\BibitemShut {NoStop}%
\bibitem [{\citenamefont {Elkady}\ \emph {et~al.}(2015)\citenamefont {Elkady},
  \citenamefont {Iskakova},\ and\ \citenamefont {Zubarev}}]{Elkady2015257}%
  \BibitemOpen
  \bibfield  {author} {\bibinfo {author} {\bibfnamefont {A.S.}\ \bibnamefont
  {Elkady}}, \bibinfo {author} {\bibfnamefont {L.}~\bibnamefont {Iskakova}}, \
  and\ \bibinfo {author} {\bibfnamefont {A.}~\bibnamefont {Zubarev}},\
  }\bibfield  {title} {\enquote {\bibinfo {title} {On the self-assembly of
  net-like nanostructures in ferrofluids},}\ }\href {\doibase
  10.1016/j.physa.2015.01.053} {\bibfield  {journal} {\bibinfo  {journal}
  {Physica A: Statistical Mechanics and its Applications}\ }\textbf {\bibinfo
  {volume} {428}},\ \bibinfo {pages} {257--265} (\bibinfo {year}
  {2015})}\BibitemShut {NoStop}%
\bibitem [{\citenamefont {Pshenichnikov}\ and\ \citenamefont
  {Ivanov}(2012)}]{Pshenichnikov2012}%
  \BibitemOpen
  \bibfield  {author} {\bibinfo {author} {\bibfnamefont {A.F.}\ \bibnamefont
  {Pshenichnikov}}\ and\ \bibinfo {author} {\bibfnamefont {A.S.}\ \bibnamefont
  {Ivanov}},\ }\bibfield  {title} {\enquote {\bibinfo {title} {Magnetophoresis
  of particles and aggregates in concentrated magnetic fluids},}\ }\href
  {\doibase 10.1103/PhysRevE.86.051401} {\bibfield  {journal} {\bibinfo
  {journal} {Physical Review E}\ }\textbf {\bibinfo {volume} {86}},\ \bibinfo
  {pages} {051401} (\bibinfo {year} {2012})}\BibitemShut {NoStop}%
\bibitem [{\citenamefont {Daff\'e}\ \emph {et~al.}(2020)\citenamefont
  {Daff\'e}, \citenamefont {Zečevi\'c}, \citenamefont {Trohidou},
  \citenamefont {Sikora}, \citenamefont {Rovezzi}, \citenamefont {Carvallo},
  \citenamefont {Vasilakaki}, \citenamefont {Neveu}, \citenamefont {Meeldijk},
  \citenamefont {Bouldi}, \citenamefont {Gavrilov}, \citenamefont {Guyodo},
  \citenamefont {Choueikani}, \citenamefont {Dupuis}, \citenamefont {Taverna},
  \citenamefont {Sainctavit},\ and\ \citenamefont {Juhin}}]{Daffe202011222}%
  \BibitemOpen
  \bibfield  {author} {\bibinfo {author} {\bibfnamefont {N.}~\bibnamefont
  {Daff\'e}}, \bibinfo {author} {\bibfnamefont {J.}~\bibnamefont {Zečevi\'c}},
  \bibinfo {author} {\bibfnamefont {K.N.}\ \bibnamefont {Trohidou}}, \bibinfo
  {author} {\bibfnamefont {M.}~\bibnamefont {Sikora}}, \bibinfo {author}
  {\bibfnamefont {M.}~\bibnamefont {Rovezzi}}, \bibinfo {author} {\bibfnamefont
  {C.}~\bibnamefont {Carvallo}}, \bibinfo {author} {\bibfnamefont
  {M.}~\bibnamefont {Vasilakaki}}, \bibinfo {author} {\bibfnamefont
  {S.}~\bibnamefont {Neveu}}, \bibinfo {author} {\bibfnamefont {J.D.}\
  \bibnamefont {Meeldijk}}, \bibinfo {author} {\bibfnamefont {N.}~\bibnamefont
  {Bouldi}}, \bibinfo {author} {\bibfnamefont {V.}~\bibnamefont {Gavrilov}},
  \bibinfo {author} {\bibfnamefont {Y.}~\bibnamefont {Guyodo}}, \bibinfo
  {author} {\bibfnamefont {F.}~\bibnamefont {Choueikani}}, \bibinfo {author}
  {\bibfnamefont {V.}~\bibnamefont {Dupuis}}, \bibinfo {author} {\bibfnamefont
  {D.}~\bibnamefont {Taverna}}, \bibinfo {author} {\bibfnamefont
  {P.}~\bibnamefont {Sainctavit}}, \ and\ \bibinfo {author} {\bibfnamefont
  {A.}~\bibnamefont {Juhin}},\ }\bibfield  {title} {\enquote {\bibinfo {title}
  {Bad neighbour, good neighbour: how magnetic dipole interactions between soft
  and hard ferrimagnetic nanoparticles affect macroscopic magnetic properties
  in ferrofluids},}\ }\href {\doibase 10.1039/d0nr02023k} {\bibfield  {journal}
  {\bibinfo  {journal} {Nanoscale}\ }\textbf {\bibinfo {volume} {12}},\
  \bibinfo {pages} {11222--11231} (\bibinfo {year} {2020})}\BibitemShut
  {NoStop}%
\bibitem [{\citenamefont {Ilg}(2017)}]{Ilg2017}%
  \BibitemOpen
  \bibfield  {author} {\bibinfo {author} {\bibfnamefont {P.}~\bibnamefont
  {Ilg}},\ }\bibfield  {title} {\enquote {\bibinfo {title} {Equilibrium
  magnetization and magnetization relaxation of multicore magnetic
  nanoparticles},}\ }\href {\doibase 10.1103/PhysRevB.95.214427} {\bibfield
  {journal} {\bibinfo  {journal} {Physical Review B}\ }\textbf {\bibinfo
  {volume} {95}},\ \bibinfo {pages} {214427} (\bibinfo {year}
  {2017})}\BibitemShut {NoStop}%
\bibitem [{\citenamefont {Pshenichnikov}\ and\ \citenamefont
  {Kuznetsov}(2015)}]{Pshenichnikov2015}%
  \BibitemOpen
  \bibfield  {author} {\bibinfo {author} {\bibfnamefont {A.F.}\ \bibnamefont
  {Pshenichnikov}}\ and\ \bibinfo {author} {\bibfnamefont {A.A.}\ \bibnamefont
  {Kuznetsov}},\ }\bibfield  {title} {\enquote {\bibinfo {title}
  {Self-organization of magnetic moments in dipolar chains with restricted
  degrees of freedom},}\ }\href {\doibase 10.1103/PhysRevE.92.042303}
  {\bibfield  {journal} {\bibinfo  {journal} {Physical Review E}\ }\textbf
  {\bibinfo {volume} {92}},\ \bibinfo {pages} {042303} (\bibinfo {year}
  {2015})}\BibitemShut {NoStop}%
\bibitem [{\citenamefont {Solovyova}\ \emph {et~al.}(2020)\citenamefont
  {Solovyova}, \citenamefont {Kuznetsov},\ and\ \citenamefont
  {Elfimova}}]{Solovyova2020}%
  \BibitemOpen
  \bibfield  {author} {\bibinfo {author} {\bibfnamefont {A.Yu.}\ \bibnamefont
  {Solovyova}}, \bibinfo {author} {\bibfnamefont {A.A.}\ \bibnamefont
  {Kuznetsov}}, \ and\ \bibinfo {author} {\bibfnamefont {E.A.}\ \bibnamefont
  {Elfimova}},\ }\bibfield  {title} {\enquote {\bibinfo {title} {Correlations
  in the simple cubic lattice of ferroparticles: Theory and computer
  simulations},}\ }\href {\doibase 10.1016/j.physa.2020.124923} {\bibfield
  {journal} {\bibinfo  {journal} {Physica A}\ }\textbf {\bibinfo {volume}
  {558}},\ \bibinfo {pages} {124923} (\bibinfo {year} {2020})}\BibitemShut
  {NoStop}%
\bibitem [{\citenamefont {Elfimova}\ \emph {et~al.}(2017)\citenamefont
  {Elfimova}, \citenamefont {Ivanov}, \citenamefont {Popescu},\ and\
  \citenamefont {Socoliuc}}]{Elfimova201754}%
  \BibitemOpen
  \bibfield  {author} {\bibinfo {author} {\bibfnamefont {E.A.}\ \bibnamefont
  {Elfimova}}, \bibinfo {author} {\bibfnamefont {A.O.}\ \bibnamefont {Ivanov}},
  \bibinfo {author} {\bibfnamefont {L.B.}\ \bibnamefont {Popescu}}, \ and\
  \bibinfo {author} {\bibfnamefont {V.}~\bibnamefont {Socoliuc}},\ }\bibfield
  {title} {\enquote {\bibinfo {title} {Transverse magneto-optical anisotropy in
  bidisperse ferrofluids with long range particle correlations},}\ }\href
  {\doibase 10.1016/j.jmmm.2016.09.051} {\bibfield  {journal} {\bibinfo
  {journal} {Journal of Magnetism and Magnetic Materials}\ }\textbf {\bibinfo
  {volume} {431}},\ \bibinfo {pages} {54--58} (\bibinfo {year}
  {2017})}\BibitemShut {NoStop}%
\bibitem [{\citenamefont {Ivanov}\ and\ \citenamefont
  {Camp}(2018)}]{Ivanov2018}%
  \BibitemOpen
  \bibfield  {author} {\bibinfo {author} {\bibfnamefont {A.O.}\ \bibnamefont
  {Ivanov}}\ and\ \bibinfo {author} {\bibfnamefont {P.J.}\ \bibnamefont
  {Camp}},\ }\bibfield  {title} {\enquote {\bibinfo {title} {Theory of the
  dynamic magnetic susceptibility of ferrofluids},}\ }\href {\doibase
  10.1103/PhysRevE.98.050602} {\bibfield  {journal} {\bibinfo  {journal}
  {Physical Review E}\ }\textbf {\bibinfo {volume} {98}},\ \bibinfo {pages}
  {050602} (\bibinfo {year} {2018})}\BibitemShut {NoStop}%
\bibitem [{\citenamefont {Solovyova}\ \emph {et~al.}(2017)\citenamefont
  {Solovyova}, \citenamefont {Elfimova}, \citenamefont {Ivanov},\ and\
  \citenamefont {Camp}}]{Solovyova2017}%
  \BibitemOpen
  \bibfield  {author} {\bibinfo {author} {\bibfnamefont {A.Y.}\ \bibnamefont
  {Solovyova}}, \bibinfo {author} {\bibfnamefont {E.A.}\ \bibnamefont
  {Elfimova}}, \bibinfo {author} {\bibfnamefont {A.O.}\ \bibnamefont {Ivanov}},
  \ and\ \bibinfo {author} {\bibfnamefont {P.J.}\ \bibnamefont {Camp}},\
  }\bibfield  {title} {\enquote {\bibinfo {title} {Modified mean-field theory
  of the magnetic properties of concentrated, high-susceptibility, polydisperse
  ferrofluids},}\ }\href {\doibase 10.1103/PhysRevE.96.052609} {\bibfield
  {journal} {\bibinfo  {journal} {Physical Review E}\ }\textbf {\bibinfo
  {volume} {96}},\ \bibinfo {pages} {052609} (\bibinfo {year}
  {2017})}\BibitemShut {NoStop}%
\bibitem [{\citenamefont {Minina}\ \emph {et~al.}(2018)\citenamefont {Minina},
  \citenamefont {Blaak},\ and\ \citenamefont {Kantorovich}}]{Minina2018}%
  \BibitemOpen
  \bibfield  {author} {\bibinfo {author} {\bibfnamefont {E.S.}\ \bibnamefont
  {Minina}}, \bibinfo {author} {\bibfnamefont {R.}~\bibnamefont {Blaak}}, \
  and\ \bibinfo {author} {\bibfnamefont {S.S.}\ \bibnamefont {Kantorovich}},\
  }\bibfield  {title} {\enquote {\bibinfo {title} {Pressure and compressibility
  factor of bidisperse magnetic fluids},}\ }\href {\doibase
  10.1088/1361-648X/aab137} {\bibfield  {journal} {\bibinfo  {journal} {Journal
  of Physics Condensed Matter}\ }\textbf {\bibinfo {volume} {30}},\ \bibinfo
  {pages} {145101} (\bibinfo {year} {2018})}\BibitemShut {NoStop}%
\bibitem [{\citenamefont {Szalai}\ \emph {et~al.}(2013)\citenamefont {Szalai},
  \citenamefont {Nagy},\ and\ \citenamefont {Dietrich}}]{Szalai2013}%
  \BibitemOpen
  \bibfield  {author} {\bibinfo {author} {\bibfnamefont {I.}~\bibnamefont
  {Szalai}}, \bibinfo {author} {\bibfnamefont {S.}~\bibnamefont {Nagy}}, \ and\
  \bibinfo {author} {\bibfnamefont {S.}~\bibnamefont {Dietrich}},\ }\bibfield
  {title} {\enquote {\bibinfo {title} {Comparison between theory and
  simulations for the magnetization and the susceptibility of polydisperse
  ferrofluids},}\ }\href {\doibase 10.1088/0953-8984/25/46/465108} {\bibfield
  {journal} {\bibinfo  {journal} {Journal of Physics Condensed Matter}\
  }\textbf {\bibinfo {volume} {25}},\ \bibinfo {pages} {465108} (\bibinfo
  {year} {2013})}\BibitemShut {NoStop}%
\bibitem [{\citenamefont {Nagornyi}\ \emph {et~al.}(2020)\citenamefont
  {Nagornyi}, \citenamefont {Socoliuc}, \citenamefont {Petrenko}, \citenamefont
  {Almasy}, \citenamefont {Ivankov}, \citenamefont {Avdeev}, \citenamefont
  {Bulavin},\ and\ \citenamefont {Vekas}}]{Nagornyi2020}%
  \BibitemOpen
  \bibfield  {author} {\bibinfo {author} {\bibfnamefont {A.V.}\ \bibnamefont
  {Nagornyi}}, \bibinfo {author} {\bibfnamefont {V.}~\bibnamefont {Socoliuc}},
  \bibinfo {author} {\bibfnamefont {V.I.}\ \bibnamefont {Petrenko}}, \bibinfo
  {author} {\bibfnamefont {L.}~\bibnamefont {Almasy}}, \bibinfo {author}
  {\bibfnamefont {O.I.}\ \bibnamefont {Ivankov}}, \bibinfo {author}
  {\bibfnamefont {M.V.}\ \bibnamefont {Avdeev}}, \bibinfo {author}
  {\bibfnamefont {L.A.}\ \bibnamefont {Bulavin}}, \ and\ \bibinfo {author}
  {\bibfnamefont {L.}~\bibnamefont {Vekas}},\ }\bibfield  {title} {\enquote
  {\bibinfo {title} {Structural characterization of concentrated aqueous
  ferrofluids},}\ }\href {\doibase 10.1016/j.jmmm.2020.166445} {\bibfield
  {journal} {\bibinfo  {journal} {Journal of Magnetism and Magnetic Materials}\
  }\textbf {\bibinfo {volume} {501}},\ \bibinfo {pages} {166445} (\bibinfo
  {year} {2020})}\BibitemShut {NoStop}%
\bibitem [{\citenamefont {Lebedev}\ \emph {et~al.}(2019)\citenamefont
  {Lebedev}, \citenamefont {Stepanov}, \citenamefont {Kuznetsov}, \citenamefont
  {Ivanov},\ and\ \citenamefont {Pshenichnikov}}]{Lebedev2019}%
  \BibitemOpen
  \bibfield  {author} {\bibinfo {author} {\bibfnamefont {A.V.}\ \bibnamefont
  {Lebedev}}, \bibinfo {author} {\bibfnamefont {V.I.}\ \bibnamefont
  {Stepanov}}, \bibinfo {author} {\bibfnamefont {A.A.}\ \bibnamefont
  {Kuznetsov}}, \bibinfo {author} {\bibfnamefont {A.O.}\ \bibnamefont
  {Ivanov}}, \ and\ \bibinfo {author} {\bibfnamefont {A.F.}\ \bibnamefont
  {Pshenichnikov}},\ }\bibfield  {title} {\enquote {\bibinfo {title} {Dynamic
  susceptibility of a concentrated ferrofluid: The role of interparticle
  interactions},}\ }\href {\doibase 10.1103/PhysRevE.100.032605} {\bibfield
  {journal} {\bibinfo  {journal} {Physical Review E}\ }\textbf {\bibinfo
  {volume} {100}},\ \bibinfo {pages} {032605} (\bibinfo {year}
  {2019})}\BibitemShut {NoStop}%
\bibitem [{\citenamefont {Linke}\ and\ \citenamefont
  {Odenbach}(2015)}]{Linke2015}%
  \BibitemOpen
  \bibfield  {author} {\bibinfo {author} {\bibfnamefont {J.M.}\ \bibnamefont
  {Linke}}\ and\ \bibinfo {author} {\bibfnamefont {S.}~\bibnamefont
  {Odenbach}},\ }\bibfield  {title} {\enquote {\bibinfo {title} {Anisotropy of
  the magnetoviscous effect in a ferrofluid with weakly interacting magnetite
  nanoparticles},}\ }\href {\doibase 10.1088/0953-8984/27/17/176001} {\bibfield
   {journal} {\bibinfo  {journal} {Journal of Physics Condensed Matter}\
  }\textbf {\bibinfo {volume} {27}},\ \bibinfo {pages} {176001} (\bibinfo
  {year} {2015})}\BibitemShut {NoStop}%
\bibitem [{\citenamefont {Pousaneh}\ and\ \citenamefont
  {De~Wijn}(2020)}]{Pousaneh2020}%
  \BibitemOpen
  \bibfield  {author} {\bibinfo {author} {\bibfnamefont {F.}~\bibnamefont
  {Pousaneh}}\ and\ \bibinfo {author} {\bibfnamefont {A.S.}\ \bibnamefont
  {De~Wijn}},\ }\bibfield  {title} {\enquote {\bibinfo {title} {Kinetic theory
  and shear viscosity of dense dipolar hard sphere liquids},}\ }\href {\doibase
  10.1103/PhysRevLett.124.218004} {\bibfield  {journal} {\bibinfo  {journal}
  {Physical Review Letters}\ }\textbf {\bibinfo {volume} {124}},\ \bibinfo
  {pages} {218004} (\bibinfo {year} {2020})}\BibitemShut {NoStop}%
\bibitem [{\citenamefont {Balescu}(1975)}]{Balescu1975}%
  \BibitemOpen
  \bibfield  {author} {\bibinfo {author} {\bibfnamefont {R.}~\bibnamefont
  {Balescu}},\ }\href@noop {} {\emph {\bibinfo {title} {Equlibrium and
  nonequlibrium statistical mechanics}}}\ (\bibinfo  {publisher} {Wiley},\
  \bibinfo {year} {1975})\BibitemShut {NoStop}%
\bibitem [{\citenamefont {Henderson}(2011)}]{Henderson2011}%
  \BibitemOpen
  \bibfield  {author} {\bibinfo {author} {\bibfnamefont {D.}~\bibnamefont
  {Henderson}},\ }\bibfield  {title} {\enquote {\bibinfo {title} {Some simple
  results for the properties of polar fluids},}\ }\href {\doibase
  10.5488/CMP14.33001} {\bibfield  {journal} {\bibinfo  {journal} {Condensed
  Matter Physics}\ }\textbf {\bibinfo {volume} {14}},\ \bibinfo {pages} {33001}
  (\bibinfo {year} {2011})}\BibitemShut {NoStop}%
\bibitem [{\citenamefont {Joslin}(1981)}]{Joslin19811507}%
  \BibitemOpen
  \bibfield  {author} {\bibinfo {author} {\bibfnamefont {C.G.}\ \bibnamefont
  {Joslin}},\ }\bibfield  {title} {\enquote {\bibinfo {title} {The third
  dielectric and pressure virial coefficients of dipolar hard sphere fluids},}\
  }\href {\doibase 10.1080/00268978100101111} {\bibfield  {journal} {\bibinfo
  {journal} {Molecular Physics}\ }\textbf {\bibinfo {volume} {42}},\ \bibinfo
  {pages} {1507--1518} (\bibinfo {year} {1981})}\BibitemShut {NoStop}%
\bibitem [{\citenamefont {Elfimova}\ \emph {et~al.}(2012)\citenamefont
  {Elfimova}, \citenamefont {Ivanov},\ and\ \citenamefont
  {Camp}}]{Elfimova2012}%
  \BibitemOpen
  \bibfield  {author} {\bibinfo {author} {\bibfnamefont {E.A.}\ \bibnamefont
  {Elfimova}}, \bibinfo {author} {\bibfnamefont {A.O.}\ \bibnamefont {Ivanov}},
  \ and\ \bibinfo {author} {\bibfnamefont {P.J.}\ \bibnamefont {Camp}},\
  }\bibfield  {title} {\enquote {\bibinfo {title} {Thermodynamics of dipolar
  hard spheres with low-to-intermediate coupling constant},}\ }\href {\doibase
  10.1103/PhysRevE.86.021126} {\bibfield  {journal} {\bibinfo  {journal}
  {Physical Review E}\ }\textbf {\bibinfo {volume} {86}},\ \bibinfo {pages}
  {021126} (\bibinfo {year} {2012})}\BibitemShut {NoStop}%
\bibitem [{\citenamefont {Elfimova}\ \emph {et~al.}(2013)\citenamefont
  {Elfimova}, \citenamefont {Ivanov},\ and\ \citenamefont
  {Camp}}]{Elfimova2013}%
  \BibitemOpen
  \bibfield  {author} {\bibinfo {author} {\bibfnamefont {E.A.}\ \bibnamefont
  {Elfimova}}, \bibinfo {author} {\bibfnamefont {A.O.}\ \bibnamefont {Ivanov}},
  \ and\ \bibinfo {author} {\bibfnamefont {P.J.}\ \bibnamefont {Camp}},\
  }\bibfield  {title} {\enquote {\bibinfo {title} {Thermodynamics of
  ferrofluids in applied magnetic fields},}\ }\href {\doibase
  10.1103/PhysRevE.88.042310} {\bibfield  {journal} {\bibinfo  {journal}
  {Physical Review E}\ }\textbf {\bibinfo {volume} {88}},\ \bibinfo {pages}
  {042310} (\bibinfo {year} {2013})}\BibitemShut {NoStop}%
\bibitem [{\citenamefont {Solovyova}\ and\ \citenamefont
  {Elfimova}(2020)}]{Solovyova2020_poly}%
  \BibitemOpen
  \bibfield  {author} {\bibinfo {author} {\bibfnamefont {A.Yu.}\ \bibnamefont
  {Solovyova}}\ and\ \bibinfo {author} {\bibfnamefont {E.A.}\ \bibnamefont
  {Elfimova}},\ }\bibfield  {title} {\enquote {\bibinfo {title} {The initial
  magnetic susceptibility of high-concentrated, polydisperse ferrofluids:
  universal theoretical expression},}\ }\href@noop {} {\bibfield  {journal}
  {\bibinfo  {journal} {Journal of Magnetism and Magnetic Materials}\ }\textbf
  {\bibinfo {volume} {495}},\ \bibinfo {pages} {165846} (\bibinfo {year}
  {2020})}\BibitemShut {NoStop}%
\bibitem [{\citenamefont {Allen}\ and\ \citenamefont
  {Tildesley}(1987)}]{Allen1987}%
  \BibitemOpen
  \bibfield  {author} {\bibinfo {author} {\bibfnamefont {M.~P.}\ \bibnamefont
  {Allen}}\ and\ \bibinfo {author} {\bibfnamefont {D.~J.}\ \bibnamefont
  {Tildesley}},\ }\href@noop {} {\emph {\bibinfo {title} {Computer simulation
  of liquids}}}\ (\bibinfo  {publisher} {Clarendon Press, Oxford},\ \bibinfo
  {year} {1987})\BibitemShut {NoStop}%
\bibitem [{\citenamefont {Kretschmer}\ and\ \citenamefont
  {Binder}(1979)}]{kretschmer1979ordering}%
  \BibitemOpen
  \bibfield  {author} {\bibinfo {author} {\bibfnamefont {R.}~\bibnamefont
  {Kretschmer}}\ and\ \bibinfo {author} {\bibfnamefont {K.}~\bibnamefont
  {Binder}},\ }\bibfield  {title} {\enquote {\bibinfo {title} {Ordering and
  phase transitions in ising systems with competing short range and dipolar
  interactions},}\ }\href {\doibase 10.1007/BF01325203} {\bibfield  {journal}
  {\bibinfo  {journal} {Zeitschrift f{\"u}r Physik B Condensed Matter}\
  }\textbf {\bibinfo {volume} {34}},\ \bibinfo {pages} {375--392} (\bibinfo
  {year} {1979})}\BibitemShut {NoStop}%
\bibitem [{\citenamefont {Budkov}\ and\ \citenamefont
  {Kiselev}(2018)}]{Budkov2018}%
  \BibitemOpen
  \bibfield  {author} {\bibinfo {author} {\bibfnamefont {Y.A.}\ \bibnamefont
  {Budkov}}\ and\ \bibinfo {author} {\bibfnamefont {M.G.}\ \bibnamefont
  {Kiselev}},\ }\bibfield  {title} {\enquote {\bibinfo {title} {Flory-type
  theories of polymer chains under different external stimuli},}\ }\href
  {\doibase 10.1088/1361-648X/aa9f56} {\bibfield  {journal} {\bibinfo
  {journal} {Journal of Physics Condensed Matter}\ }\textbf {\bibinfo {volume}
  {30}},\ \bibinfo {pages} {043001} (\bibinfo {year} {2018})}\BibitemShut
  {NoStop}%
\bibitem [{\citenamefont {Budkov}\ and\ \citenamefont
  {Kolesnikov}(2016)}]{Budkov2016}%
  \BibitemOpen
  \bibfield  {author} {\bibinfo {author} {\bibfnamefont {Yu.A.}\ \bibnamefont
  {Budkov}}\ and\ \bibinfo {author} {\bibfnamefont {A.L.}\ \bibnamefont
  {Kolesnikov}},\ }\bibfield  {title} {\enquote {\bibinfo {title} {On a new
  application of the path integrals in polymer statistical physics},}\ }\href
  {\doibase 10.1088/1742-5468/2016/10/103211} {\bibfield  {journal} {\bibinfo
  {journal} {Journal of Statistical Mechanics: Theory and Experiment}\ }\textbf
  {\bibinfo {volume} {2016}},\ \bibinfo {pages} {103211} (\bibinfo {year}
  {2016})}\BibitemShut {NoStop}%
\end{thebibliography}%

\end{document}